\documentclass[11pt,a4paper]{article}
\usepackage{jheppub}
\pdfoutput=1
\title{Electron and muon  anomalous magnetic dipole moment in a 3-3-1 model }

\author[a]{G. De Conto}
\author[a]{and V. Pleitez}

\affiliation[a]{
Instituto  de F\'\i sica Te\'orica--Universidade Estadual Paulista \\
R. Dr. Bento Teobaldo Ferraz 271, Barra Funda\\ S\~ao Paulo - SP, 01140-070,
Brazil
}

\emailAdd{georgedc@ift.unesp.br}
\emailAdd{vicente@ift.unesp.br}

\begin{document}
		
\maketitle
\flushbottom

\begin{abstract}

We calculate, in the context of a 3-3-1 model with heavy charged leptons, constraints on some parameters of the extra particles in the model by imposing that their contributions to both the electron and muon $(g-2)$ factors are in agreement with experimental data up to 1$\sigma$-3$\sigma$. In order to obtain realistic results we use some of the possible solutions of the left- and right- unitary matrices that diagonalize the lepton mass matrices,  giving the observed lepton masses and at the same time allowing to accommodate the Pontecorvo-Maki-Nakagawa-Sakata (PMNS) mixing matrix. 
We show that, at least up to 1-loop order, in the particular range of the space parameter that we have explored, it is not possible to fit the observed electron and muon $(g-2)$ factors at the same time unless one of the extra leptons has a mass of the order of 20-40 GeVs and the energy scale of the 331 symmetry to be of around 60-80 TeVs.  
 	
\end{abstract}

\section{Introduction}

Both anomalous magnetic dipole moments (AMDM) of electron and muon~$a_{e,\mu}~=~(g~-~2)_{e,\mu}/2$, have been measured with great precision. We  have $a^{\textrm{exp}}_e= 1159652180.76 (0.28) \times 10^{-12} $ for the electron, and $a^{\textrm{exp}}_\mu = 11659209(6) \times 10^{-10} $ for the muon~\cite{Agashe:2014kda}. The theoretical calculations within the Standard Model (SM) have reached the high level of precision of the experiments. On one hand, the calculation for the electron $a_e$, considering only QED up to tenth order, gives the result of $a^{\textrm{SM}}_e= 1 159 652 181.643\times  10^{-12}$~\cite{Aoyama:2014sxa}, giving a difference of
\begin{equation}
\Delta a_e =\mu^{\textrm{exp}}_e-\mu^{\textrm{SM}}_e=-1.05(82)\times 10^{-12},
\label{mdme}
\end{equation} 
which is close to one standard deviation. On the other hand, for the muon $a_\mu$, calculations considering QED up to five loops and hadronic vacuum polarization up to next-to-leading order, hadronic light-by-light scattering and electroweak contributions results in $a^{SM}_\mu~=~1.16591801(49)~\times 10^{-3}$~\cite{Jegerlehner:2009ry,Miller:2012opa,Blum:2013xva}, leading to a difference between theory and experiment of
\begin{equation} 
\Delta a_\mu=\mu^{\textrm{exp}}_\mu-\mu^{\textrm{SM}}_\mu=2.87(80)\times 10^{-9},
\label{mdmmu}
\end{equation}
a difference that goes beyond three standard deviations from the experimental result.

This difference between the SM prediction and the experimental value of the muon anomalous magnetic moment has been studied in many models. 
For instance, in other 3-3-1 models~\cite{Kelso:2013zfa,Kelso:2014qka,Ky:2000ku,Binh:2015cba}, left-right symmetric models~\cite{Taibi:2015ura}, supersymmetric models~\cite{Chowdhury:2015rja,Harigaya:2015jba,Ajaib:2015yma,Padley:2015uma,Khalil:2015wua} and two-Higgs doublets models~\cite{Abe:2015oca,Chun}. 

In particular, the 3-3-1 models are interesting extensions of the standard model (SM), since they solve some of the questions that the SM  leaves without answer. For example, among others, the number of generations, why $\sin^2\theta_W<1/4$, and electric charge quantization. The models have also interesting consequences in flavor physics~\cite{Buras:2012dp,Buras:2014yna,Buras:2016qia} and, in particular, those 3-3-1 models with quarks with electric charge -4/3 and 5/3 (in units of $\vert e \vert$) have at least one neutral scalar which can appear as a heavy resonance that could be observed at LHC~\cite{Martinez:2015kmn,Hernandez:2015ywg,Cao:2015scs,Martinez:2016ztt}. The model has new singly and doubly charged vector bosons that can also appear as LHC resonances~\cite{deBlas:2015hlv}. %In fact, if the existence of this resonance is confirmed in the near future, one natural explanation for it is the minimal 3-3-1 model. We use here \textit{natural} in the sense that usually models are proposed just to solve particular discrepancies with the SM, as the latter ones. Meanwhile, 3-3-1 models assume a new set of gauge symmetries and explore their consequences, supposing that this new set of symmetries may explain all experimental results. However, we stress that it is still early to say if that resonance is a physical effect or a statistical fluctuation.

At the same time that the agreement between the experiment and the SM for the electron AMDM  gives us confidence in the correctness of the theory, the disagreement in the muon case suggests that there may be effects unaccounted by the SM, or it is still possible that such effects come from new particles and through their interactions with the already known particles. For instance, the contributions of heavy leptons to the $a_\mu$ factor have been considered in gauge models since Refs.~\cite{Bjorken:1972am,Primack:1973iz}. However, we should not forget that if the AMDM of the muon really implies new physics, its value should disagree with the SM beyond 5$\sigma$. 

In the literature, when the explanation of the observed value of the muon $a_\mu$ is based on new physics, its effects in the electron $a_e$ are usually not taken into account. It is explicitly assumed that these effects do not perturb the values of the electron AMDM. Here we will show that, at least in this particular model, it is not possible to fit the $\Delta a_e$ and $\Delta a_\mu$ with the same parameters considering 1-loop order calculations, unless one of the extra leplons has a mass of the order of 20-40 GeV and the energy scale of the 331 symmetry is of tens of TeVs. Also, usually in literature the charged lepton masses are neglected, thus allowing to obtain simple expressions for the $\Delta a_\mu$. In the latter cases, $\Delta a_\mu(X)\propto m^2_\mu/M^2_X$. 
Although this case seems to be reasonable, it is a quite optimistic one. See Sec.~\ref{sec:con} for a more detailed discussion. 

Here we will consider the contributions to both $a_{e,\mu}$ in the context of the 3-3-1 model with heavy leptons (331HL for short). For more details of the model see Refs.~\cite{Pleitez:1992xh} and \cite{DeConto:2014fza}. 
An important issue, that in general cannot be neglected in 3-3-1 models, are the contributions of the scalars and pseudo-scalars, since sometimes there are important positive and/or negative interference with the amplitudes involving other particles in the model~\cite{Machado:2013jca}. However, in the 331HL that we are considering, since there are no flavor violating neutral currents in the leptonic sector, these vertices are proportional to the lepton masses and thus they are negligible. This is because, unlike the minimal 3-3-1 model, there is only one source for all lepton masses. 

The outline of the paper is as follows. In the next section we review briefly only the lepton and scalar sectors of the  model. In Sec.~\ref{sec:mdm} we show general expressions of each of the contributions for the MDM of a given charged lepton, which we identified with the electron or the muon. In Sec.~\ref{sec:results} we give the parameters that are used in the calculations with the interactions given in the Appendix \ref{sec:interactions}, and the last section is devoted to our conclusions. 
In Appendix \ref{sec:potential} we show the scalar potential. 

\section{The 3-3-1 model}
\label{sec:model}

Models with gauge symmetry $SU(3)_C\otimes SU(3)_L\otimes U(1)_X$ present new possibilities for the electroweak interactions. Here we consider the 3-3-1 model with heavy charged leptons (331HL for short) in which there are new exotic quarks and leptons. Moreover, to give mass to all the particles, more scalar fields are needed. Hence, these models are intrinsically multi-Higgs models. 
In this model the electric charge operator is given by
$Q/|e|=T_{3}-\sqrt{3}T_{8}+X$,
where $e$ is the electron charge, $T_{3,8}=\lambda_{3,8}/2$ (being $\lambda_{3,8}$ the Gell-Mann matrices) and $X$ is the hypercharge operator associated to the $U(1)_X$ group. Below, since the quarks and gauge bosons are the same as in the minimal 3-3-1 model, we only present the scalar and lepton content of the model, with its charges associated to each group on the parentheses, in the form ($SU(3)_C$, $SU(3)_L$, $U(1)_X$).

The minimal scalar sector for the model is composed by three triplets \cite{Pleitez:1992xh}:\\$\chi~=~(\chi^{-}  \chi^{--} \chi^{0})^T\sim(1,3,-1)$, $\rho=(\rho^{+} \rho^{0} \rho^{++})^T \sim(1,3,1)$
and $\eta=(\eta^{0} \eta_{1}^{-} \eta_{2}^{+})^T~\sim~(1,3,0)$,
where $\chi^0=\frac{|v_\chi|e^{i\theta_\chi}}{\sqrt{2}}\left( 1+\frac{X^0_\chi+i I^0_\chi}{|v_\chi|} \right)$ and 
$\psi^0=\frac{|v_\psi|}{\sqrt{2}}\left( 1+\frac{X^0_\psi+i I^0_\psi}{|v_\chi|} \right)$, for $\psi=\,\eta,\,\rho$. Some versions of the minimal 331 model, with less triplets than usual, are in trouble with phenomenology~\cite{Montero:2011tg}, thus we use the three scalar triplets above. The leptonic sector have three left handed triplets and six right handed singlets:
$\Psi_{aL}=(\nu^\prime_{a} \, l^\prime_{a} \, E^\prime_{a})_L^T\sim (1,3,0)$ and the respective right-handed singlets $\nu^\prime_{aR}\sim(1,1,0)$, $l^\prime_{aR} \sim (1,1,-1)$, and $E^\prime_{aR}\sim(1,1,1)$, where $a=e,\mu,\tau$. 
Here $\nu^\prime_a,l^\prime_a$ denote the usual leptons and $E^\prime_a$ new leptons with positive electric charge. 
Primed fields denotes symmetry eigenstates.

The Yukawa Lagrangian in the lepton sector is given by:
\begin{equation}
-\mathcal{L}_Y^l= G^\nu_{ab}\bar{\Psi}_{aL}\nu^\prime_{bR}\eta+G^l_{ab}\bar{\Psi}_{aL}l^{'-}_{bR}\rho + G^E_{ab}\bar{\Psi}_{aL}E^{\prime+}_{bR}\chi+M_{ab}\overline{(\nu^\prime_{aR})^c}\nu^\prime_{bR}+H.c.
\label{yukawa}
\end{equation}
where $G^\nu,G^l$ and $G^E$ are arbitrary $3\times3$ matrices. From (\ref{yukawa}), we obtain the Yukawa interactions given in Appendix~\ref{sec:interactions} and the mass matrices for the leptons, which are given by  
$M^l=G^l|v_\rho|/\sqrt{2}$ for the $l^\prime$-type leptons and $M^E=G^E |v_\chi|e^{i\theta_\chi}/\sqrt{2}$ for the $E^\prime$-type leptons (the neutrinos are massless at this stage). For simplicity we are assuming the $G^E$ matrix to be diagonal. In this manner, for our masses to be real, we need that the elements of the $G^E$ matrix to have the form $(G^E)_{ii}=|G^E|_{ii}e^{-i\theta_\chi}$, which implies $m_{E_i}=\delta_{ii}G^E_i|v_\chi|/\sqrt{2}$ 
(where $E_1=E_e,E_2=E_\mu,E_3=E_\tau$). Notice from Eq.~(\ref{yukawa}) that in the 331HL model there are no flavor changing neutral currents mediated by scalar fields. Notice that neutrinos have all the ingredients for a type I seesaw mechanism. Moreover, we note that the interaction $\bar{E}^\prime_{aL}\nu^\prime_{bR}\eta^+_2$ does exist and, as we will show later, it is important to make the extra leptons $E^\prime$ unstable. 

The mass eigenstates for the non-exotic leptons are obtained as $l^\prime_{L,R}=(V_{L,R}^l)^\dagger l_{L,R}$, where $l^\prime=(l_1,l_2,l_3)$, and $l=(e,\mu,\tau)$ (the neutrinos symmetry eigenstates corresponds to the mass eigenstates). 
These $V_{L,R}^{l}$ matrices diagonalize the mass matrix in the following manner:  $V^l_L M^l V^{l\dagger}_R=diag(m_e,m_\mu,m_\tau)$.
Possible solutions for the $V_{L,R}^{l}$ matrices and the $G^l$ matrix can be found in \cite{DeConto:2014fza} and \cite{Machado:2016jzb}.
To find these solutions it was considered $|v_\rho|=54$ GeV and $|v_\eta|=240$ GeV, as in Ref.~\cite{Machado:2013jca}, these will be used in our analysis of the muon and electron AMDMs. Since the masses of the exotic leptons are unknown, we cannot find such solution for them and for this reason, as we said above, we considered their mass matrix diagonal. 
Although neutrinos get mass, say by the type-I seesaw mechanism~\cite{DeConto:2015eia}, because of the small neutrino masses, the effect of unitary matrices in the vertices involving singly charged scalars, for all practical processes, is negligible in its non-diagonal elements: they are suppressed by the small neutrino masses.     The neutrino masses are not of direct interest for the calculation of the AMDM thus, we only note that if $V^\nu_L$, is the matrix which diagonalize the active neutrino masses, we can define the PMNS matrix as $V_{PMNS}=V^{l\dagger}_LV^\nu_L$.  Then, it is possible that accommodate both $PMNS$ and the active neutrino masses as it was done in Refs.~\cite{DeConto:2015eia,Machado:2016jzb}.

\section{MDM in the 331HL model}
\label{sec:mdm}

\begin{figure}
\includegraphics[width=\textwidth]{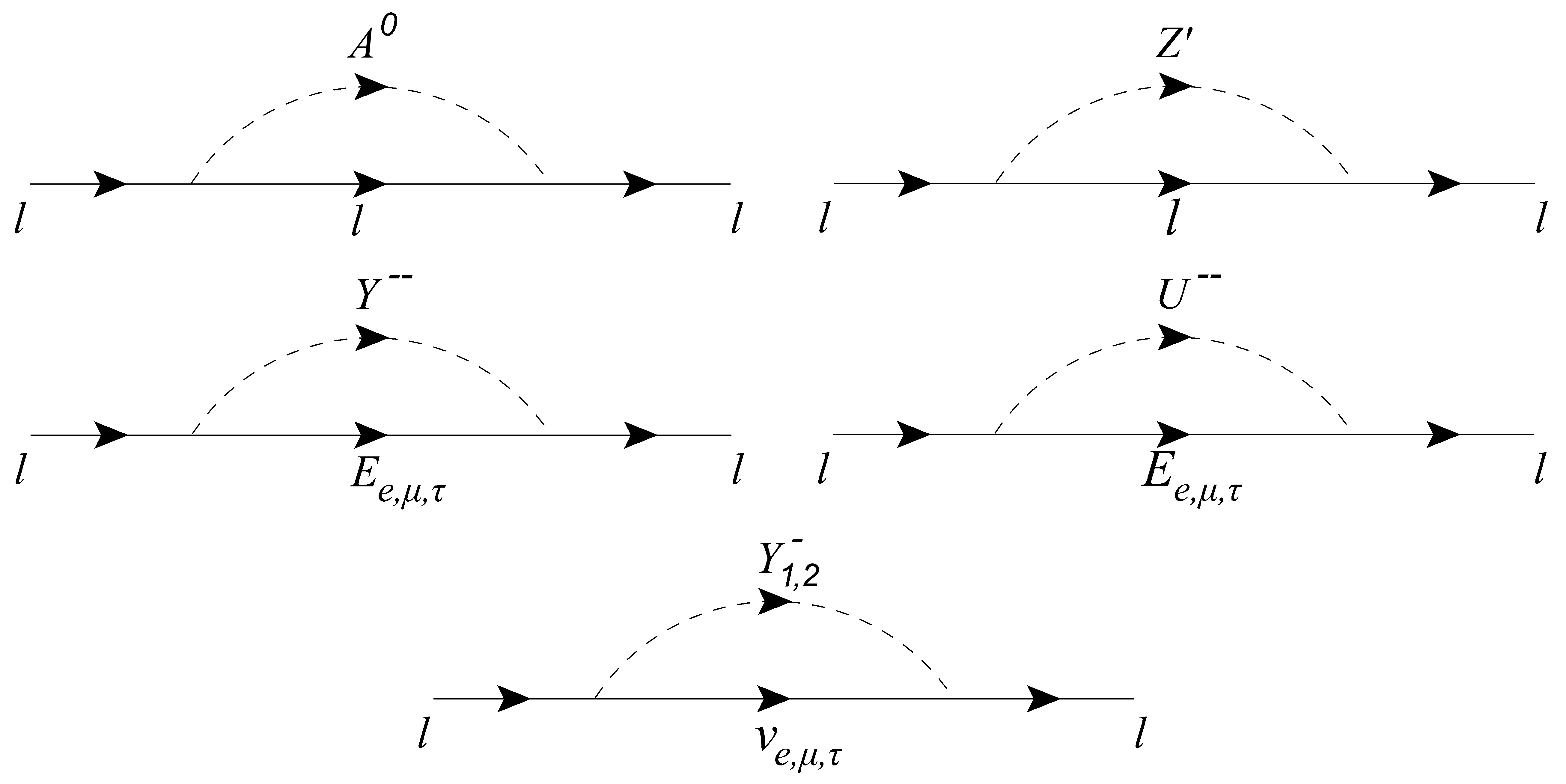}
\caption{1 loop diagrams for the MDM. The fermion $l$ indicates either an electron or a muon. The 3-3-1 model contribution to the MDM comes from all these diagrams, considering two cases, when the photon is connected to the fermion and when the photon is connected to the boson (when applicable).}
\label{fig:MDM_todos}
\end{figure}

In the 331HL model, the main extra contributions to MDMs arise from the heavy leptons, several scalars and vector bileptons (see Fig.~\ref{fig:MDM_todos}). Here we present the one-loop contributions due to these  particles in the model. We consider the scalar-lepton vertexes to have the form $i(S_\xi+P_\xi\gamma_5)$  where $\xi=h_i,A^0,Y^+_1,Y^{++}$, with the factors $S_\xi,P_\xi$ given in Eqs.~(\ref{aae2})-(\ref{aae4}). The vector-lepton vertices are considered to have the form $i\gamma^\mu(V_U-A_U\gamma_5)$ for the vector $U^{--}$ and $i\gamma^\mu(f_V-f_A\gamma_5)$ for the vector $Z^\prime$. All these couplings are given in Eqs.~(\ref{aae10})-(\ref{aae14}). We present the general result below, valid either for the electron or for the muon (the diagrams were calculated in the  unitary gauge). The full result comes from considering all the possible one-loop diagrams involving any exotic lepton, scalar or vector particles. We will verify if only the extra contributions in the 331HL model are enough to satisfy the constraints in Eqs.~(\ref{mdme}) and (\ref{mdmmu}). In this case, all the considered diagrams are those shown in Fig.~\ref{fig:MDM_todos}. The muon  MDM has been considered in the context of other 3-3-1 models in \cite{Kelso:2013zfa,Kelso:2014qka,Ky:2000ku}, however, there the authors did not consider the lepton mixing and also the constraints coming from the electron $(g-2)_e$. As can be seen in the matrices shown in Secs. \ref{sec:NumericalResultsAMDMsub1}-\ref{sec:NumericalResultsAMDMsub4}, it is possible that there are important non-diagonal entries in the matrix $V^l_L$. Moreover, solving the muon $\Delta a _\mu$ discrepancy and at the same time giving contributions compatible with $\Delta a_e$ is not a trivial issue, at least in 1-loop order.

As said before, we consider only the extra diagrams present in the model as being responsible for the new contributions to the $\Delta a_\mu$, such that
\begin{equation}
a_i^{331}=\Delta a_i,\quad i=e,\mu,
\label{oba1}
\end{equation}
where $a^{331}_i$ includes diagrams with at least one of the extra particles in the model. They are shown in Fig.~\ref{fig:MDM_todos}. 

\subsection{Scalar contributions}
\label{subsec:mdmscalars}

Unlike the minimal 331 model, in the present one there are no FCNC through the Higgs exchange. In this situation a neutral scalar $S$ has only scalar interactions $\bar{f}fS$, and a pseudo-scalar $A$ has only pseudo-scalar interactions $\bar{f}\gamma_5fA$. We will denote the respective factors $S^l_{\xi(\zeta)}$ and $P^l_{\xi(\zeta)}$ where $l$ denote the external lepton and $\xi$, or $\zeta$, denotes the scalar in the loop. 

First, we consider the case in which there is a neutral or charged scalar, denoted by $\xi$, in the loop. 
When the photon is connected to the fermion ($f$) line we have:
\begin{equation}
\Delta a^l_{\xi}(f)=-\frac{Q_I}{96\pi^2}m^2_l\sum_\xi\frac{1}{M^2_{\xi}}\int_0^1 dx \frac{\vert P^l_\xi\vert^2 F^\xi_P(x,\epsilon^\xi_l)+\vert S^l_\xi\vert^2F^\xi_S(x,\epsilon^\xi_l)}
{F^\xi(x,\epsilon^\xi_l,\lambda^\xi_l)},
\label{as}
\end{equation}
with $\epsilon^\xi_l=m_I/m_l$, $\lambda^\xi_l=m_l/M_\xi$, and $\xi=h_1,A,Y^-_1,Y^{--}$, $m_l$ is the mass of the electron or muon, and $M_\xi$ is the scalar mass in the loop;  $S^l_\xi$ and $P^l_\xi$ are the matrices given in Eq.~(\ref{aae2}) and depend on the type of the scalar, $Q_I(m_I)$ is the electric charge (mass) of the internal lepton. We have defined
\begin{eqnarray}
&& F^\xi_S(x,\epsilon^\xi_l)=-g(x)+12 \epsilon^\xi_l (x+1),\nonumber \\&&
F^\xi_P(x,\epsilon^\xi_l)=g(x)+12 \epsilon^\xi_l (x+1),\nonumber \\ && 
F^\xi(x,\epsilon^\xi_l,\lambda^\xi_l)=x[(\lambda^{\xi}_l)^2(x-1)+1]-(\epsilon^\xi_l\lambda^\xi_l)^2(x-1),
\label{ff1}
\end{eqnarray} 
and $\epsilon^{h,A}_l=1$  (again, this is because there are no flavor changing neutral currents via neutral scalar or pseudo-scalars) and $\epsilon^{Y^{--}}_l=m_I/m_l$, with $M_I=E_e,E_\mu,E_\tau$, and $\epsilon^{Y^{-}_1}_l=m_\nu/m_l$; we have also defined $g(x)=12x^2+9x-1$. Moreover, $\epsilon^A_l\lambda^A_l=m_l/M_A$ (similarly for $h$), and $\epsilon^{Y^{--}}_l\lambda^{Y^{--}}_l=m_{E_l}/M_{Y^{--}}$. 

In the case when the boson is a  singly or  doubly charged scalar, there are also diagrams in which the photon is connected to the scalar line. In the case of $Y^{--}$  we obtain 
\begin{equation}
\Delta a^l_\zeta(\zeta)=\frac{Q_\zeta}{8\pi^2} m^2_l \sum_\zeta\frac{1}{M^2_\zeta}\int_0^1 dx \,
\frac{\vert P^l_\zeta \vert^2R^\zeta_1(x,\epsilon^\zeta_l)+\vert S^l_\zeta\vert^2R^\zeta_2(x,\epsilon^\zeta_l)}{R^\zeta(x,\epsilon^\zeta_l,\lambda^\zeta_l)  },
\end{equation}
for the scalar couplings.  When $\zeta=Y^{--}$ we define 
\begin{eqnarray}
&& R^{Y^{--}}_1 (x,\epsilon^{Y^{--}}_l)=x[  -(1-x)+\epsilon^{Y^{--}}_l],\nonumber \\ &&
R^{Y^{--}}_2(x,\epsilon^{Y^{--}}_l)=x[1-x+\epsilon^{Y^{--}}_l], \nonumber \\&&  R^{Y^{--}}(x,\epsilon^{Y^{--}}_l,\lambda^{Y^{--}}_l)=
x[\lambda^2_l(x-1) +\left(\epsilon^{Y^{--}}_l\lambda^{Y^{--}}_l\right)^2]- (x-1),
\label{ff2}
\end{eqnarray}
and when $\zeta=Y^-_{1,2}$ we have
\begin{eqnarray}
&& R^{Y^-_{1,2}}_1 (x,\epsilon^-_l)=-x(1-x),\nonumber \\ &&
R^{Y^-_{1,2}}_2(x,\epsilon^{Y^-_{1,2}}_l)=x(1-x), \nonumber \\&&  R^{Y^-_{1,2}}(x,\epsilon^{Y^-_{1,2}}_l,\lambda^{Y^-_{1,2}}_l)=
(\lambda^{Y^-_{1,2}}_l)^2(x-1)^2,
\label{ff3}
\end{eqnarray}
where we have neglected $\epsilon^{Y^-_1}_\nu=m_\nu/m_l$ for both electron and muon. In fact, we are neglecting the PMNS matrix and assuming for practical purposes that the $\nu$-$Y^-_1$-$l$ vertex is diagonal. This means that we are over-estimating the contributions of the $Y^-_{1,2}$ scalar but even in this case we will see that their contributions are negligible. 

\subsection{Vector contributions}
\label{subsec:gaugefields}

The model has neutral and doubly charge vector bosons that contribute to the AMDM in one loop, $Z^\prime_\mu$ and $U^{--}_\mu$, respectively. In the latter case, when the photon is connected to the charged  vector line, we have:
\begin{equation}
\Delta a^l_U(U)=-\frac{G_{UUA}}{64\pi^2}\frac{m^2_l}{M^2_U}  \int_0^1 dx \left[\frac{T_1(x,\epsilon^U_l)+T_2(x,\epsilon^U_l)+T_3(x,\epsilon^U_l)
T(x,\epsilon^U_l,\lambda^U_l)}{T(x,\epsilon^U_l,\lambda^U_l)}
\right],
\label{mu1}
\end{equation}
with $\lambda^U_l=m_l/M_U$, where $M_U$ is the mass of the $U^{--}_\mu(E)$; $\epsilon^U_l=m_I/m_l,\;E_I=E_e,E_\mu,E_\tau$. We have both vector and axial-vector couplings:
\begin{eqnarray}
&& T_1(x,\epsilon^u_l)=  \vert A^l_U\vert^2[h(x)+3 \epsilon^U_l (x+1)]
+ \vert V^l_U\vert^2[h(x)+3\epsilon^U_l(x+1)], \nonumber \\ &&
T_2(x,\epsilon^U_l)= \frac{m^2_l}{M^2_U}x^3[\vert A^l_U\vert^2 (x+2 \epsilon^U_l)+
\vert V^l_U\vert^2(2\epsilon^U_l-x)],\nonumber \\&&
T_3(x,\epsilon^U_l)= 2\vert A^l_U\vert^2[h(x)+4+\epsilon^U_l(2x+1)]
+\vert V^l_U\vert^2[-(2x+1)^2+\epsilon^U_l(6x+1)], \nonumber \\ &&
T(x,\epsilon^U_l,\lambda_l)=  x-1-x [(\lambda^U_l)^2 (x-1)+(\epsilon^U_l\lambda^U_l)^2],
\label{tes1}
\end{eqnarray}
where we have defined $h(x)=2x^2+x-3$.  

The factor $G_{UUA}=-2e$, see Eq. (\ref{aae9}), and $A^l_U$ and $V^l_U$  are given in Sec.~\ref{subsec:glint}. There is only one diagram of this type, that with $U^{--}$ in Fig.~\ref{fig:MDM_todos}. 

For the case where the photon is connected to the fermion line there are two diagrams in 
Fig.~\ref{fig:MDM_todos}, one with $Z^\prime$ and the other with $U^{--}$:
\begin{equation}
\Delta a^l_X(f)=-\frac{Q_I}{8\pi^2}m^2_l\sum_X \frac{1}{M^2_X} \int_0^1 dx\left[\frac{\tilde{R}^X_1(x,\epsilon^X_l)+\tilde{R}^X_2(x,\epsilon^X_l)+
\tilde{R}^X_3(x,\epsilon^X_l)
\tilde{R}^X(x,\epsilon^X_l,\lambda^X_l)}{\tilde{R}^X(x,\epsilon^X_l,\lambda^X_l)}
\right],
\label{mu2}
\end{equation}
where $X=Z^\prime_\mu,U^{--}_\mu$, and$M$ is the mass of the vector boson, $Q_I$ is the electric charge of the fermion internal in the loops, $\lambda^X_l=m_l/M_X $, with 
\begin{eqnarray}
&& \tilde{R}^X_1(x,\epsilon^X_l)= 2\vert A^l_X\vert^2 (1+x+2 \epsilon^X_l)+2\vert V^l_X\vert^2(1+x-2\epsilon^X_l),
\nonumber \\ && 
\tilde{R}^X_2(x,\epsilon^X_l)= \frac{m^2_l}{M^2_X} (x-1)^2\left[\vert A^l_X\vert^2(1+\epsilon^X_l) 
(x-\epsilon^X_l )+\vert V^l_X\vert^2(1-\epsilon^X_l)(x+\epsilon^X_l)\right],
\nonumber \\ && 
\tilde{R}^X_3(x,\epsilon^X_l)=  (3 x-1)[\vert A^l_X\vert^2(1+\epsilon^X_l)+\vert V^l_X\vert^2(1-\epsilon^X_l)],\nonumber \\ &&
\tilde{R}^X(x,\epsilon^X_l,\lambda^X_l)=1+(x-1) [(\lambda^X_l)^2+(\epsilon^X_l\lambda^X_l)^2].
\label{res1}
\end{eqnarray}

In (\ref{tes1}) and (\ref{res1}), when $X=U^{--}_\mu$ the matrices $A^l_U$ and $V^l_U$ are those in Sec.~\ref{subsec:glint} and, when $X=Z^\prime_\mu$, $A^l_{Z^\prime}$ and $V^l_{Z^\prime}$ are given by the factors $f^l_A$ and $f^l_V$ in Eq.~(\ref{aae14}).

Notice that when the vector is the $Z^\prime$, $\epsilon^{Z^\prime}_l=1$. But with $U^{--}_\mu$ defined as above: $\epsilon^U_l=m_E/m_l$.
We have assumed $M_X\gg m_I$ but no approximation was made.  For all the vertices the reader is referred to Appendices~\ref{subsec:leptonscalars} - \ref{subsec:geugeleptoninteractions}.

\section{Results for the electron and the muon MDM}
\label{sec:results}

Considering the results in the previous section we are able to find sets of values for the parameters of the model taking into account the contributions of the scalars $A^0,Y_1^-,Y^{++}$ and the vector bosons $V^-_\mu,U^{--}_\mu$ for the electron and the muon MDM, that matches the difference between the SM predictions and the experimental results for both cases in (\ref{mdme}) and (\ref{mdmmu}). Let us define the contributions of the extra particles in the 331HL as in (\ref{oba1}), where $\Delta a_{e,\mu}$ are given in (\ref{mdme}) and (\ref{mdmmu}), respectively, in such a way to obtain $\mu_{331}+\mu_{SM}=\mu_{exp}$ within 1 to 3 standard deviations. Before we show our results, let us consider what sort of heavy leptons the $E_l$ are in the present model.

\subsection{Decays of the extra charged leptons}
\label{subsec:decays}

The experimental searches of new charged leptons have considered three scenarios in order to put constraints on the masses of this sort of particles~\cite{Agashe:2014kda}: i) sequential heavy leptons, in which these particles are assumed as belonging to a fourth generation where either the neutrino partner is considered stable, or the heavy leptons decay into active neutrinos via mixing; ii) stable heavy charged leptons; and iii) long lived heavy charged leptons. In situation i) the lower limit on the mass of such particle is 100.8 GeV; in ii) it is 102.6 GeV, and finally in case iii) it is 574 GeV. In the  present model the heavy leptons $E^\pm$ are not stable but may be long-lived depending on the their masses and on the masses of the scalar bosons that mediate their decays.
In particular, we note that the lower limit on the mass of a sequential heavy charged leptons, 100.8 GeV at 95\%, was obtained using the decay $\ell^- \to W^-\nu_l$~\cite{Agashe:2014kda}. However, in the present model, as we will show below, the decays are mediated mainly by extra charged vector and scalar bosons and, for this reason, it is not straightforward to apply this limit on the masses of $E^\pm$ in the present model. Besides, in the 331HL there are three of such leptons and only one of them can be constrained by the experimental searches.

The lepton triplets in terms of the symmetry eigenstates (primed fields) are $\Psi_{aL}=(\nu^\prime_{a} \, l^\prime_{a} \, E^\prime_{a})^T\sim (1,3,0)$, $a=e,\mu,\tau$~\cite{Pleitez:1992xh}. Usually it is assumed that the leptons $E^+_a$ are antiparticles and thus $L(E^+_a)=-1$. In this case the $L$ assignment is 
\begin{equation}
L(J,j,\eta^-_2,\chi^-,\chi^{--},\rho^{--},V^-_\mu,U^{--}_\mu)=+2.
\label{exotics}
\end{equation}
and the other particles having $L=0$, or $+1$. In this case the model has a global custodial symmetry $U(1)_L$ under which some particles (including the usual ones) are $L=1$ (antiparticles $L=-1$) and the other ones as in Eq.~(\ref{exotics}). We can also use the global symmetry $U(1)_{F}$, where 
$F=B+L$~\cite{Pleitez:1993gc}. 

However, we can assume that the leptons $E^+_a$ are particles, \textit{\`a la} Konopinski-Mahmoud~\cite{Konopinski:1953gq}, and assign it $L(E^+_a)=+1$. In this case the lepton number is the same for all members in lepton triplets, and $L(J,j,\eta^-_2,\chi^-,\chi^{--},\rho^{--},V^-_\mu,U^{--}_\mu)=0$. Notwithstanding, a custodial discrete $\mathbb{Z}_2$ symmetry still exists, under which $E,J,j,\eta^-_2,\chi^-,\chi^{--},\rho^{--},V^-_\mu,U^{--}_\mu$ are odd, and the rest of the particles are even. The electroweak symmetries of the model is $G_W=\mathbb{Z}_2\times SU(3)_L\otimes U(1)_X$. In this section we call "exotic"  the particles that are odd under $\mathbb{Z}_2$, otherwise they are "normal" particles. Now, $\mathbb{Z}_2$ is the custodial symmetry.

At first sight, the custodial discrete symmetry implies that the lightest exotic charged lepton should be stable. 
There are interactions that produce $E^+\to V^+\nu_L$ and $E^+\to U^{++}l^-$ but both vectors, $V^+$ and $U^{++}$, cannot decay only into know quarks or leptons, $U^{++}\to E^+e^+$ and $V^+\to e^+_L\nu^c_R$. Moreover, these vector bosons can also decay into one exotic quark and one known quark, $V^+\to \bar{u}J$ and $U^{++}\to u\bar{j}$. We can see from Eq.~(\ref{yukawa}), that the interactions with charged scalars are similar, they involve also one normal and one exotic particle: $E^+_L\to \eta^+_2\nu_R$ and $E^+_R\to \nu_L\chi^+$, which  are allowed since $\eta^+_2$ and $\chi^+$ are odd under $\mathbb{Z}_2$. However, $\eta^+_2$ and $\chi^+$ decays also into one normal particle and one exotic one. Hence, the lightest $E^+$ cannot, at first sight, decay at all.  

However, we note that the quartic term $a_{10}(\chi^\dagger \eta ) (\rho^\dagger \eta )$ is allowed in the scalar potential and it implies a mixing among all the singly charged scalars~\cite{Montero:1998yw} (See appendix \ref{sec:potential}):  
\begin{equation}
V(\eta,\rho,\chi)\supset a_{10}(\chi^+\eta^0+\chi^{++}\eta^-_1+\chi^{0*}\eta^+_2)(\rho^-\eta^0+\rho^{0*}\eta^-_1+\rho^{--}\eta^+_2).
\end{equation}

\begin{figure}
	\begin{center}
\includegraphics[width=5.0cm]{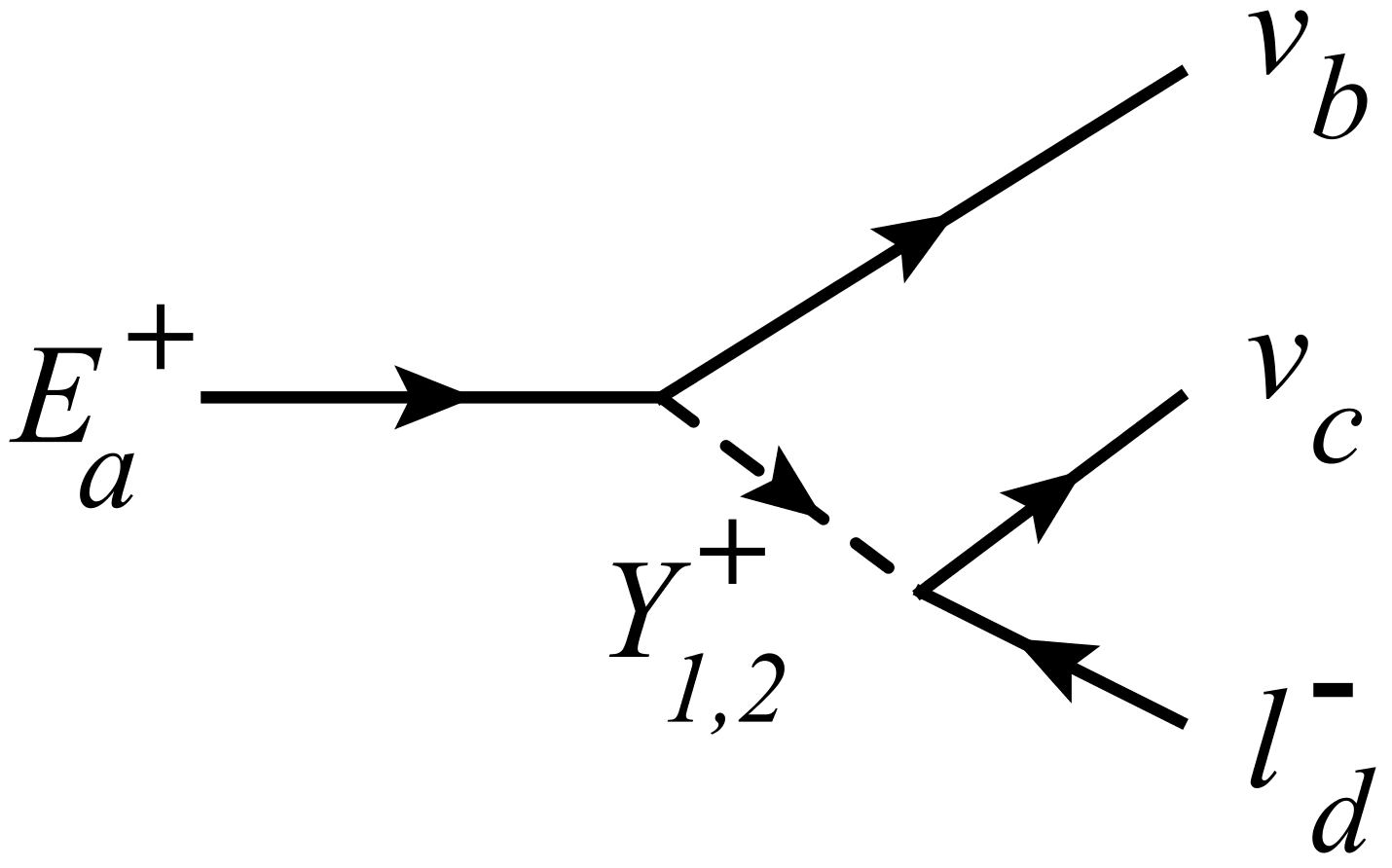}
\caption{This decay is allowed even for the lightest lepton $E$.}
\label{decayE}
	\end{center}
\end{figure}

Notice that the term  $\chi^+\eta^0\rho^-\eta^0$ breaks the $\mathbb{Z}_2$ symmetry.
Moreover, if $a_{10}\not=0$, all the singly charged scalars mix in the mass matrix and since the interaction $\bar{\nu}_LE_R\chi^-$ does exist, see Eq.~(\ref{salva}), the decay $E^+_{lR}\to \nu_{lL}+h^+\to \nu_{lL}+l^{+}+\nu^c_{l}$ is now allowed, where $h^+$ is a charged scalar mass eigenstates that couples with the known leptons (we have omitted the matrix element that projects $\chi^+$ onto $h^+$). The decay of any $E$ through charged scalars is shown in Fig.~\ref{decayE}. A rough calculation of the decay, based on the result for the muon decay $\mu \rightarrow 2\nu +e$, is given by
\begin{equation}
\tau_E= \frac{m_Y^4}{(C m_E)^4} \frac{12 (8 \pi)^3}{m_E} \hbar,
\label{vidamedia}
\end{equation}
where $m_Y$ is the mass of the lightest singly charged scalar, and $C$ denotes the couplings between the $Y$ scalar and the fermions. We see that with the reasonable values: $C=10^{-1}$, $m_E=20$ GeV and $m_Y=500$ GeV, we obtain $\tau_E\sim 2.4\times 10^{-11}$ s. If we assumed the values for the masses used in our plots in Sec. \ref{sec:NumericalResultsAMDM}, we would get even shorter lifetimes for the exotic leptons. The channel into one charged lepton and two active neutrinos is open even for the lightest of the leptons $E^+_a$ and thus it is not long lived anymore. The issue of the stability of exotic fermions deserves a more detailed study that will be published elsewhere. However, the above discussion leaves clear that the usual experimental data are not applicable, at least in a straightforward way, in the present model.

\subsection{Numerical results for the AMDMs}
\label{sec:NumericalResultsAMDM}

Here we will present the numerical results of the AMDMs for both, muon and electron, showing which parts of the parameter region satisfy the experimental results. We will consider five different scenarios, each with a different set of diagonalization matrices for the leptonic sector (see Appendix \ref{subsec:leptonscalars} for more details).
When the masses of the exotic particles are not explicitly mentioned in each plot, it means that they were fixed as: $m_{E_e}=m_{E_\mu}=m_{E_\tau}=500$ GeV, $m_{Y_2^-}= 1200$ GeV and $m_{A^0}=m_{Y^{--}}=m_{Y_1^-}=1000$ GeV. Also, the couplings for the singly charged scalars in the $\bar{\nu}EY_1^+$ and $\bar{\nu}EY_2^+$ vertices were assumed to be both 0.5, see Eq.~(\ref{aae3 }). We have tried several values for the scalar couplings, but they have given no noticeable change in our plots, that is because the $U^{++}$ contribution dominates (see Fig. \ref{fig:MDM_UPP}).
The masses of the gauge bosons ($U^{\pm\pm}$ and $Z'$) have their values defined by the value of $|v_\chi|$, since its other parameters are already fixed (see \cite{Dias:2006ns} for details). According to \cite{Agashe:2014kda}, the lowest lower limit on the mass of the $Z^\prime$ boson is 2.59 TeV, assuming it has the same couplings as the Z boson, which implies $|v_\chi| > 665.13$ GeV~\cite{Machado:2013jca}.  Here we will show that in the present model a stronger lower limit for  $v_\chi$  is obtained and that, at least with the values of the matrices $V^l_L$ used, it is not possible to fit both (\ref{mdme}) and (\ref{mdmmu}) at the same time within 1$\sigma$ at the one loop order and if all extra leptons should be heavy with masses larger that 100 GeV and if the 331 symmetry is realized up to some few TeVs. In fact, we show that it is $a_e$ which imposes a higher constraint on $v_\chi$.
 
We must remind the reader that the $C\!P$ violating phase in $v_\chi$, in the form $\cos\theta_\chi$, is present in some of the vertices, however, in a previous work considering the 331HL model~\cite{DeConto:2014fza}, we found that such phase should be no greater than $10^{-6}$, therefore we are considering in this work, for the sake of simplicity, $\theta_\chi$ to be zero.
 
 \subsubsection{Diagonal $V^l_{L,R}$ matrices}
  \label{sec:NumericalResultsAMDMsubDiag}
 
\begin{figure}
	\includegraphics[width=\textwidth]{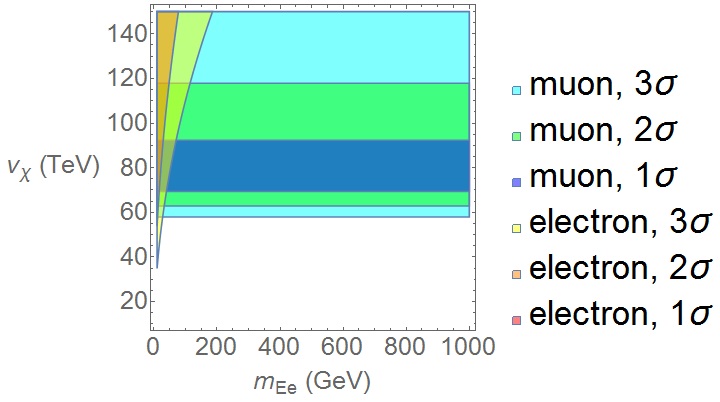} %./GraficoMDM_mEe
	\caption{$v_\chi$ and $m_{E_e}$ values satisfying Eq. (\ref{oba1}) within 1$\sigma$, 2$\sigma$ and 3$\sigma$. Here we have used diagonal mass matrices for the charged leptons (see Sec. \ref{sec:NumericalResultsAMDMsubDiag}) and considered: $m_{E_\mu}=m_{E_\tau}=$500, $m_{Y_1^+}=$1000, and  $m_{A^0}=m_{Y^{++}}=m_{Y_1^+}=$1000 (all in GeV). The masses of the exotic gauge bosons ($U^{\pm\pm}$ and $Z'$) have their values defined by the value of $v_\chi$, since its other parameters are already fixed (see \cite{Dias:2006ns} for details).}
	\label{fig:GraficoMDM_mEe_MatDiag}
\end{figure}

\begin{figure}
	\includegraphics[width=\textwidth]{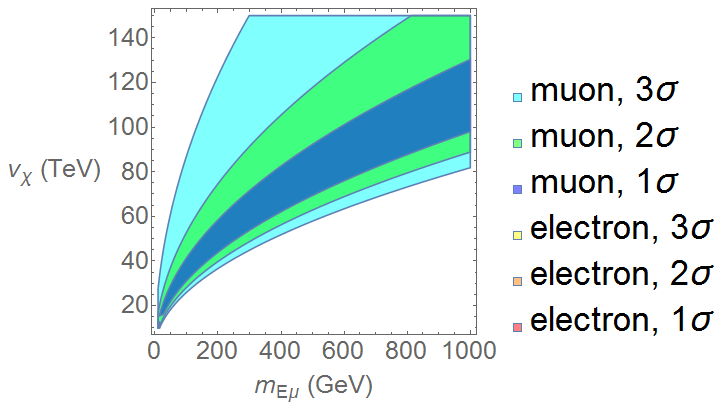} % ./GraficoMDM_mEmu
	\caption{Same as in Fig.~\ref{fig:GraficoMDM_mEe_MatDiag} but now with $m_{E_e}=m_{E_\tau}=$500 GeV.}
	\label{fig:GraficoMDM_mEmu_MatDiag}
\end{figure}

\begin{figure}
	\includegraphics[width=\textwidth]{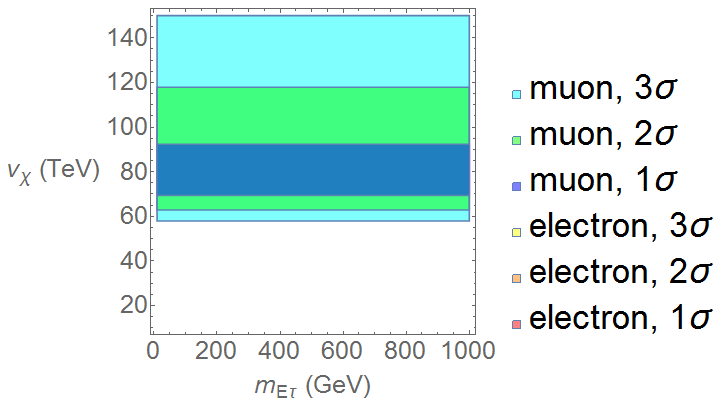} %./GraficoMDM_mEtau
	\caption{Same as in Fig.~\ref{fig:GraficoMDM_mEe_MatDiag} but now with $m_{E_e}=m_{E_\mu}=$500 GeV.}
	\label{fig:GraficoMDM_mEtau_MatDiag}
\end{figure}

 The simplest solution possible is to assume that the leptonic interactions are diagonal in flavor, i.e., the symmetry ans mass eigenstates are the same. In Fig.~\ref{fig:GraficoMDM_mEe_MatDiag} we vary the mass of the exotic lepton $E_e$, fixing the masses of the other two. In a similar manner, in figure \ref{fig:GraficoMDM_mEmu_MatDiag} we vary the mass of $E_\mu$ and in figure \ref{fig:GraficoMDM_mEtau_MatDiag} we vary the mass of $E_\tau$, fixing the other masses. The blue, green and cyan regions show values for the parameters where the 331HL-only contributions for the muon MDM agrees with the difference between the experimental results and the SM prediction within $1\sigma$ - $3\sigma$, respectively. In a similar manner, the red, orange and yellow regions show values where the 331HL-only contribution for the electron MDM agrees within $1\sigma$ and $3\sigma$, respectively. It can be seen in the figure that there are solutions for the muon up to $1\sigma$ and for the electron up to $2\sigma$. However the 1$\sigma$ and 2$\sigma$ regions only overlap for low values of $m_{E_e}$, less than 40 GeV, and $v_\chi$ around 80 TeV. These are not unrealistic solutions because an extra charged lepton lighter than 40 GeV may still exist depending of the respective couplings with the known particles, and $v_\chi$ may have a value of 60-80 TeV such that $Z^\prime$ and bileptons $V,U$ will be only of a  few tens of TeVs (See Eq. \ref{uvmasses}). Although this model has a Landau-like pole at an energy of around 4 TeV~\cite{Dias:2004dc} this only indicates the energy at which the model loses its perturbative nature. This situation does not change too much when realistic values for the matrices $V^l_L$ are considered.
 
 It is expected that high values for the exotic leptons and bosons masses would lead to results similar to the ones found in the SM, where the electron MDM deviates from the experimental results by 1.3$\sigma$. It can be seen in Fig. \ref{fig:GraficoMDM_mEe_MatDiag} that, for values of $v_\chi$ over 60 TeV and $m_{Ee}\lesssim$ 1 GeV, we have a 2$\sigma$ region for the electron MDM. Such high values for $v_\chi$ implies masses of tens of TeV's for the exotic gauge bosons, suppressing several of the diagrams shown in Fig. \ref{fig:MDM_todos}, leaving us diagrams with exotic leptons and scalars. However, for the exotic lepton masses, we only explored values up to 1 TeV, and close to this upper value we have found no solutions for the electron MDM. But, comparing with the values explored for the gauge boson masses, we expect that the diagrams containing an $E$ lepton will only be suppressed when masses of the order of tens of TeV's are considered for such particles. As for the scalar masses, they are all assumed to be around 1 TeV, also small if compared with the exotic gauge boson masses considered.
 
 We can use realistic unitary matrices which satisfy the condition $V^l_L M^l V^{l\dagger}_R=\textrm{diag}(m_e\,m_\mu\,m_\tau)$. Below we will consider four types of parametrization of these unitary matrices.
%Given that exotic leptons with such low masses are unlikely to be found (and experimentally excluded from some scenarios) it is hard to say that the 331HL can explain the muon and electron AMDMs in this case. 

%As for Figs. %\ref{fig:GraficoMDM_mEmu_MatDiag} and \ref{fig:GraficoMDM_mEtau_MatDiag}, they only show solutions for the muon AMDM.
 
 \subsubsection{1st set of $V^l_{L,R}$ matrices}
 \label{sec:NumericalResultsAMDMsub1}

The diagonalization matrices used in this set are:
\begin{equation}
V_L^l=
\left(
\begin{array}{ccc}
0.009854320681804862 & 0.31848228260886335 & -0.9478775912680647 \\
0.014570561834801654 & -0.947868712966038 & -0.3183278211340082 \\
-0.9998452835772734 & -0.010674204623999706 & -0.013981068053858256
\end{array}
\right),
\end{equation}

\begin{equation}
V_R^l=
\left(
\begin{array}{ccc}
0.005014143494893113 & 0.0026147097108665555 & 0.9999840107012414 \\
0.0071578125624917055 & 0.9999708696847197 & -0.0026505662235414198 \\
0.9999618113129783 & -0.0071709884334755225 & -0.004995281829296536 \\
\end{array}
\right).
\end{equation}

\begin{figure}
\includegraphics[width=\textwidth]{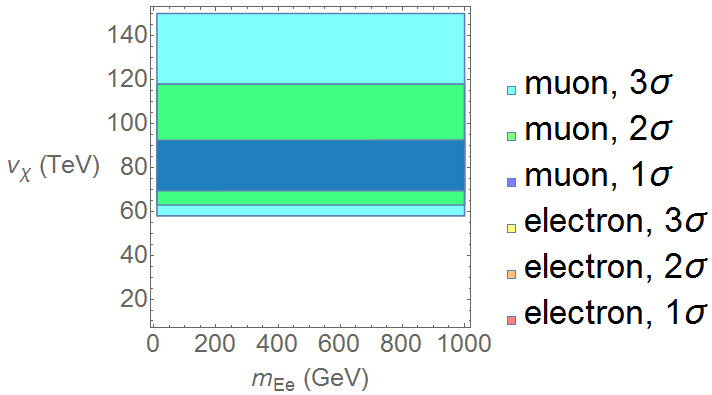} %./GraficoMDM_mEe
\caption{$v_\chi$ and $m_{E_e}$ values satisfying Eq. (\ref{oba1}) within 1$\sigma$, 2$\sigma$ and 3$\sigma$. Here we used the first set of diagonalization matrices (see Sec. \ref{sec:NumericalResultsAMDMsub1}) and considered: $m_{E_\mu}=m_{E_\tau}=$500, $m_{Y_1^+}=$1000, and  $m_{A^0}=m_{Y^{++}}=m_{Y_1^+}=$1000 (all in GeV). The masses of the exotic gauge bosons ($U^{\pm\pm}$ and $Z'$) have their values defined by the value of $v_\chi$, since its other parameters are already fixed (see \cite{Dias:2006ns} for details).}
\label{fig:GraficoMDM_mEe_MatV1}
\end{figure}

\begin{figure}
\includegraphics[width=\textwidth]{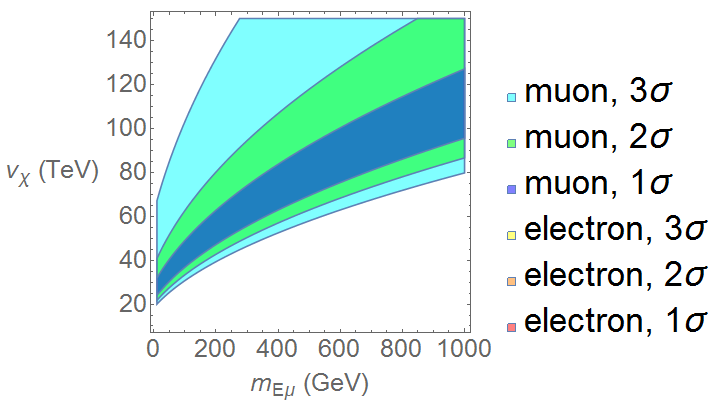} % ./GraficoMDM_mEmu
\caption{Same as in Fig.~\ref{fig:GraficoMDM_mEe_MatV1} but now with $m_{E_e}=m_{E_\tau}=$500 GeV.}
\label{fig:GraficoMDM_mEmu_matV1}
\end{figure}

\begin{figure}
\includegraphics[width=\textwidth]{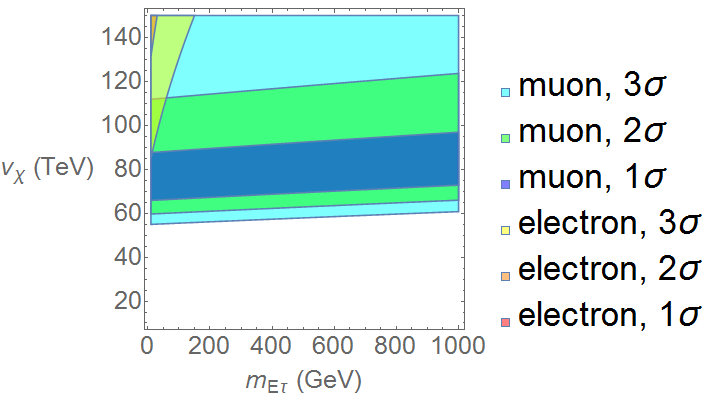} %./GraficoMDM_mEtau
\caption{Same as in Fig.~\ref{fig:GraficoMDM_mEe_MatV1} but now with $m_{E_e}=m_{E_\mu}=$500 GeV.}
\label{fig:GraficoMDM_mEtau_MatV1}
\end{figure}

Figs. \ref{fig:GraficoMDM_mEe_MatV1} and \ref{fig:GraficoMDM_mEmu_matV1} shows us only solutions for the muon. Meanwhile, in Fig. \ref{fig:GraficoMDM_mEtau_MatV1} we have solutions for the electron up to 2$\sigma$. These solutions intersect with the 3$\sigma$ solutions for the muon for values of $m_{E_\tau}$ smaller than 20 GeV and $v_\chi$ greater than 140 TeV. As for the 3$\sigma$ solutions for the electron, they intersect the muon 2$\sigma$ region for $m_{E_\tau}<$ 50 GeV and $v_\chi \sim$ 90 TeV.

\subsubsection{2nd set of $V^l_{L,R}$ matrices}
\label{sec:NumericalResultsAMDMsub2}

The diagonalization matrices used in this set are:
\begin{equation}
V_L^l=
\left(
\begin{array}{ccc}
-0.009 & 0.0146 & -0.9998 \\
-0.3185 & -0.9479 & -0.0107 \\
0.9479 & -0.3183 & -0.0140
\end{array}
\right),
\end{equation}

\begin{equation}
V_R^l=
\left(
\begin{array}{ccc}
0.005 & 0.0072 & 0.9999 \\
0.0026 & 0.9910 & -0.0072 \\
0.9999 & -0.0027 & -0.0050 \\
\end{array}
\right).
\end{equation}

\begin{figure}
	\includegraphics[width=\textwidth]{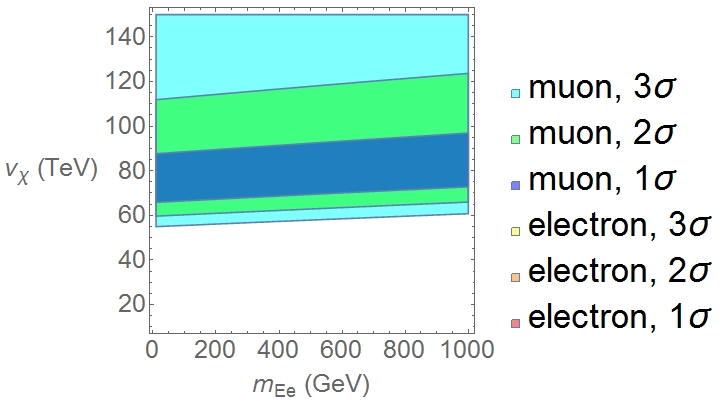} %./GraficoMDM_mEe
	\caption{$v_\chi$ and $m_{E_e}$ values satisfying Eq. (\ref{oba1}) within 1$\sigma$, 2$\sigma$ and 3$\sigma$. Here we used the second set of diagonalization matrices (see Sec. \ref{sec:NumericalResultsAMDMsub2}) and considered: $m_{E_\mu}=m_{E_\tau}=$500, $m_{Y_1^+}=$1000, and  $m_{A^0}=m_{Y^{++}}=m_{Y_1^+}=$1000 (all in GeV). The masses of the exotic gauge bosons ($U^{\pm\pm}$ and $Z'$) have their values defined by the value of $v_\chi$, since its other parameters are already fixed (see \cite{Dias:2006ns} for details).}
	\label{fig:GraficoMDM_mEe_MatV2}
\end{figure}

\begin{figure}
	\includegraphics[width=\textwidth]{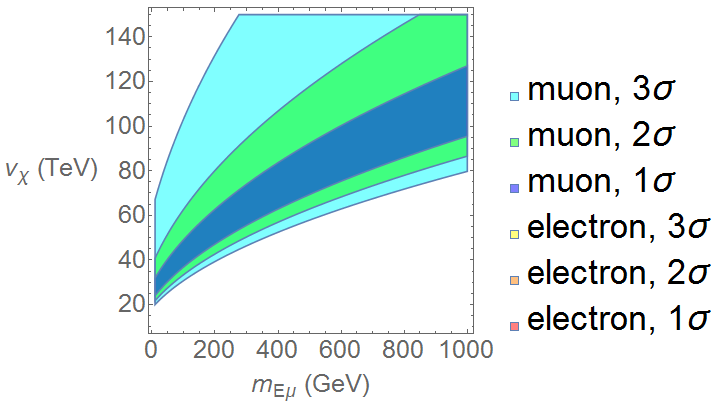} % ./GraficoMDM_mEmu
	\caption{Same as in Fig.~\ref{fig:GraficoMDM_mEe_MatV2} but now with $m_{E_e}=m_{E_\tau}=$500 GeV.}
	\label{fig:GraficoMDM_mEmu_matV2}
\end{figure}

\begin{figure}
	\includegraphics[width=\textwidth]{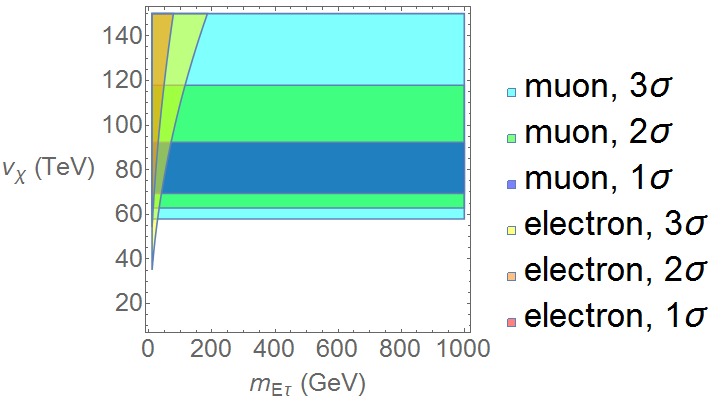} %./GraficoMDM_mEtau
	\caption{Same as in Fig.~\ref{fig:GraficoMDM_mEe_MatV2} but now with $m_{E_e}=m_{E_\mu}=$500 GeV.}
	\label{fig:GraficoMDM_mEtau_MatV2}
\end{figure}

Similar to the first set of matrices, the plots where we vary $m_{E_e}$ and $m_{E_\mu}$ (Figs. \ref{fig:GraficoMDM_mEe_MatV2} and \ref{fig:GraficoMDM_mEmu_matV2}) show only muon solutions, while in Fig. \ref{fig:GraficoMDM_mEtau_MatV2} have solutions for both electron and muon that overlap. The situation is similar to the 1st set of matrices, now with a broader range of values for $v_\chi$. Better than before, now the 2$\sigma$ electron region and the 1$\sigma$ muon region are overlapping, for values of $m_{E_\tau}\lesssim$ 15 GeV and $v_\chi \sim$80 TeV.

\subsubsection{3rd set of $V^l_{L,R}$ matrices}
\label{sec:NumericalResultsAMDMsub3}

The diagonalization matrices used in this set are:
\begin{equation}
V_L^l=
\left(
\begin{array}{ccc}
0.983908 & 0.156151 & 0.086891 \\
0.0777852 & 0.061974 & -0.994965 \\
-0.160853 & 0.985709 & 0.0500342
\end{array}
\right),
\end{equation}

\begin{equation}
V_R^l=
\left(
\begin{array}{ccc}
0.978756 & 0.186555 & 0.0850542 \\
0.0744144 & 0.0633254 & -0.99215 \\
-0.191048 & 0.980401 & 0.0480978
\end{array}
\right).
\end{equation}

\begin{figure}
	\includegraphics[width=\textwidth]{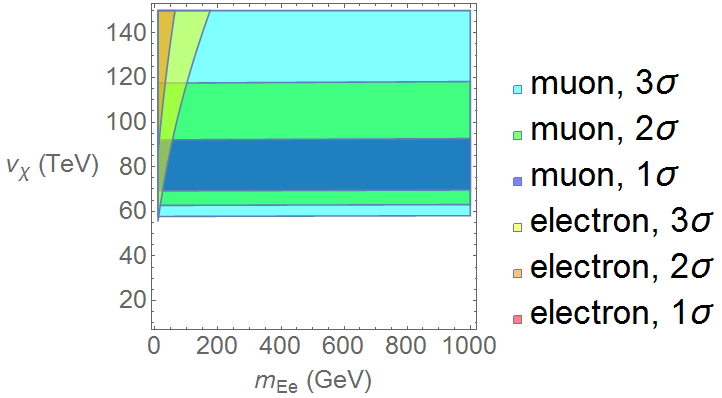} %./GraficoMDM_mEe
	\caption{$v_\chi$ and $m_{E_e}$ values satisfying Eq. (\ref{oba1}) within 1$\sigma$, 2$\sigma$ and 3$\sigma$. Here we used the third set of diagonalization matrices (see Sec. \ref{sec:NumericalResultsAMDMsub3}) and considered: $m_{E_\mu}=m_{E_\tau}=$500, $m_{Y_1^+}=$1000, and  $m_{A^0}=m_{Y^{++}}=m_{Y_1^+}=$1000 (all in GeV). The masses of the exotic gauge bosons ($U^{\pm\pm}$ and $Z'$) have their values defined by the value of $v_\chi$, since its other parameters are already fixed (see \cite{Dias:2006ns} for details).}
	\label{fig:GraficoMDM_mEe_MatV3}
\end{figure}

\begin{figure}
	\includegraphics[width=\textwidth]{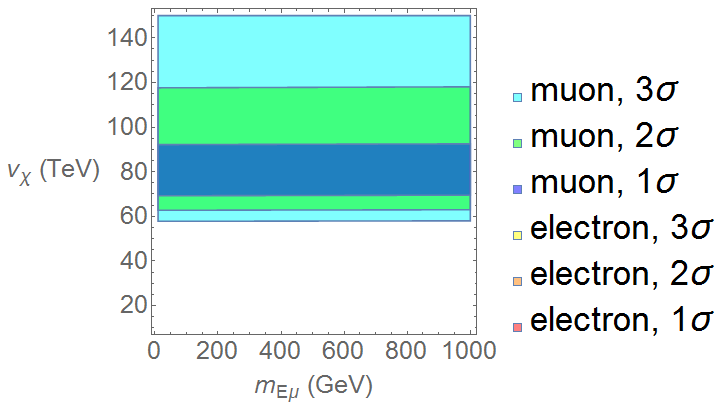} % ./GraficoMDM_mEmu
	\caption{Same as in Fig.~\ref{fig:GraficoMDM_mEe_MatV3} but now with $m_{E_e}=m_{E_\tau}=$500 GeV.}
	\label{fig:GraficoMDM_mEmu_matV3}
\end{figure}

\begin{figure}
	\includegraphics[width=\textwidth]{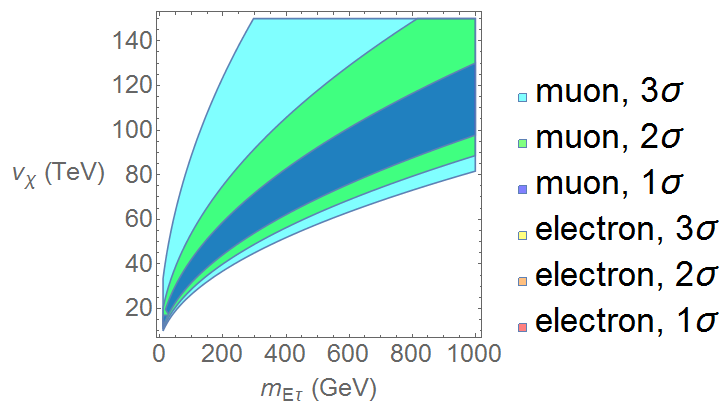} %./GraficoMDM_mEtau
	\caption{Same as in Fig.~\ref{fig:GraficoMDM_mEe_MatV3} but now with $m_{E_e}=m_{E_\mu}=$500 GeV.}
	\label{fig:GraficoMDM_mEtau_MatV3}
\end{figure}

In this set only Fig. \ref{fig:GraficoMDM_mEe_MatV3} has solutions for the muon and the electron, while figures \ref{fig:GraficoMDM_mEmu_matV3} and \ref{fig:GraficoMDM_mEtau_MatV3} have only muon solutions. The 2$\sigma$ regions for the electron and muon overlap for $m_{E_e}\lesssim$ 30 GeV and $v_\chi \sim$ 110 TeV. Also, there is an overlap of the electron 3$\sigma$ and the muon 1$\sigma$ regions, for $m_{E_e}\lesssim$ 50 GeV and $v_\chi \sim$ 80 TeV.

\subsubsection{4th set of $V^l_{L,R}$ matrices}
\label{sec:NumericalResultsAMDMsub4}

The diagonalization matrices used in this set are:

\begin{equation}
V_L^l=
\left(
\begin{array}{ccc}
-0.99614 & -0.08739 & -0.00826 \\
0.01357 & 0.24625 & -0.96691 \\
0.08672 & 0.96526 & 0.24649
\end{array}
\right),
\end{equation}

\begin{equation}
V_R^l=
\left(
\begin{array}{ccc}
0.99624 & -0.08629 & -0.00801 \\
0.01179 & 0.226594 & -0.97392 \\
0.08586 & 0.97016 & 0.22676
\end{array}
\right).
\end{equation}

\begin{figure}
	\includegraphics[width=\textwidth]{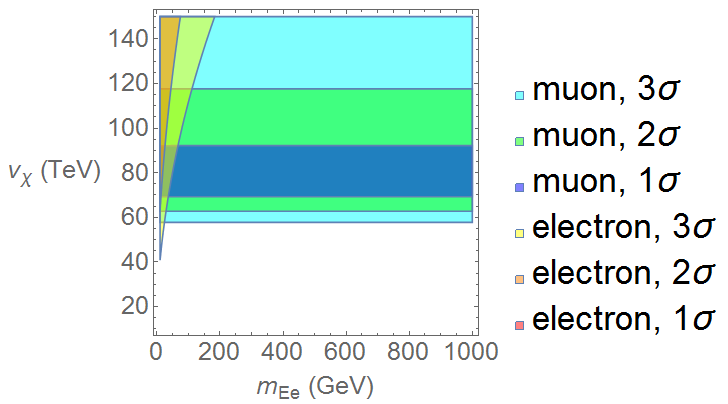} %./GraficoMDM_mEe
	\caption{$v_\chi$ and $m_{E_e}$ values satisfying Eq. (\ref{oba1}) within 1$\sigma$, 2$\sigma$ and 3$\sigma$. Here we used the fourth set of diagonalization matrices (see Sec. \ref{sec:NumericalResultsAMDMsub4}) and considered: $m_{E_\mu}=m_{E_\tau}=$500, $m_{Y_1^+}=$1000, and  $m_{A^0}=m_{Y^{++}}=m_{Y_1^+}=$1000 (all in GeV). The masses of the exotic gauge bosons ($U^{\pm\pm}$ and $Z'$) have their values defined by the value of $v_\chi$, since its other parameters are already fixed (see \cite{Dias:2006ns} for details).}
	\label{fig:GraficoMDM_mEe_MatV4}
\end{figure}

\begin{figure}
	\includegraphics[width=\textwidth]{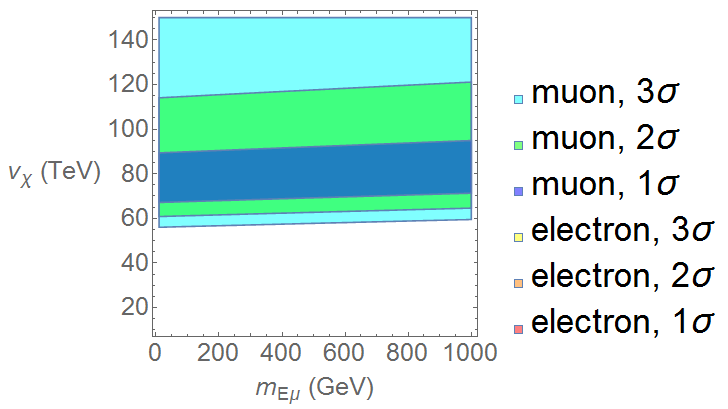} % ./GraficoMDM_mEmu
	\caption{Same as in Fig.~\ref{fig:GraficoMDM_mEe_MatV4} but now with $m_{E_e}=m_{E_\tau}=$500 GeV.}
	\label{fig:GraficoMDM_mEmu_matV4}
\end{figure}

\begin{figure}
	\includegraphics[width=\textwidth]{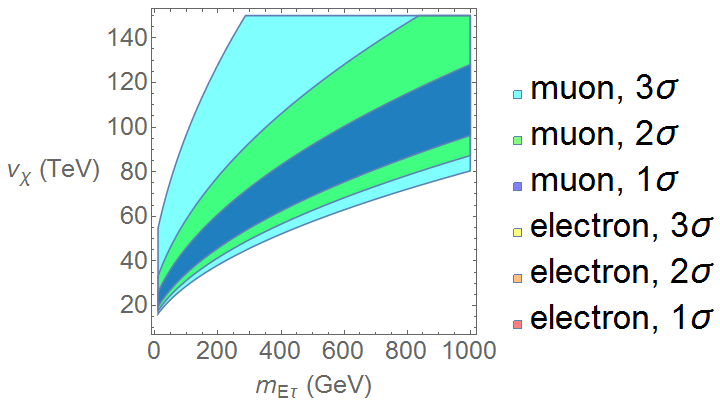} %./GraficoMDM_mEtau
	\caption{Same as in Fig.~\ref{fig:GraficoMDM_mEe_MatV4} but now with $m_{E_e}=m_{E_\mu}=$500 GeV.}
	\label{fig:GraficoMDM_mEtau_MatV4}
\end{figure}

In Fig. \ref{fig:GraficoMDM_mEe_MatV4} we see solutions for the muon and electron that overlap for low values of $m_{E_e}$, while in Figs. \ref{fig:GraficoMDM_mEmu_matV4} and \ref{fig:GraficoMDM_mEtau_MatV4} we see only solutions for the muon. The situation in Fig. \ref{fig:GraficoMDM_mEe_MatV4} is similar to that in Fig. \ref{fig:GraficoMDM_mEtau_MatV2}, where the 2$\sigma$ electron region and the 1$\sigma$ muon region are overlapping, for values of $m_{E_\tau}\lesssim$ 15 GeV and $v_\chi\sim$80 TeV.

\begin{figure}
	\includegraphics[width=\textwidth]{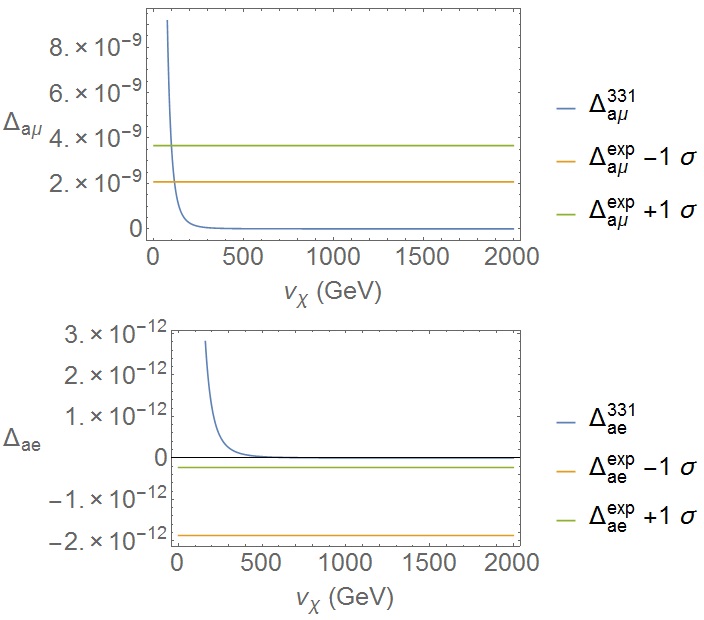}%{mY++}
	\caption{$\Delta a_e$ (upper graphic) and $\Delta a_\mu$ (lower graphic) taking into account only the contribution of the doubly charged scalar $Y^{--}$ as function of $v_\chi$, with $0< v_\chi\leq 2$ TeV, $m_{Y^{--}}=500$ GeV. We also assume that both $V_{L,R}$ are the unit matrix and $m_{Ee}=m_{E\mu}=m_{E\tau}=20$ GeV. We see that the muon MDM only has solutions for $v_\chi \approx 100$ GeV, while the electron MDM has no solutions for this value.}
	\label{fig:MDM_YPP}
\end{figure}

\begin{figure}
	\includegraphics[width=\textwidth]{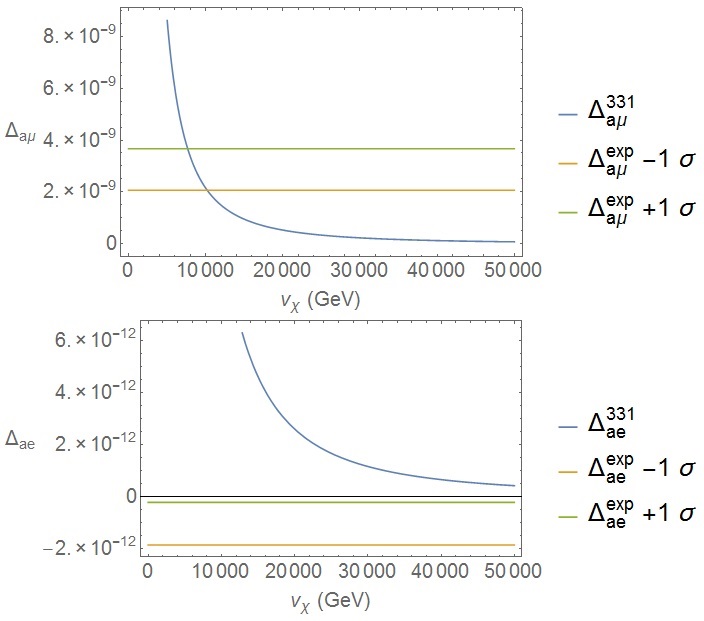}%{mU++}
	\caption{$\Delta a_e$ (upper graphic) and $\Delta a_\mu$ (lower graphic) taking into account only the contribution of the doubly charged vector boson $U^{--}$ as function of $v_\chi$, with $0< v_\chi\leq 50$ TeV. We also assume that both $V_{L,R}$ are the unit matrix and $m_{Ee}=m_{E\mu}=m_{E\tau}=20$ GeV. It can be seen that the muon MDM is solved for $v_\chi$ around 8 TeV, a value that does not solve the electron MDM.}
	\label{fig:MDM_UPP}
\end{figure}

Although we considered all the contributions to the $\Delta a_{e,\mu}$ in 1-loop in all these scenarios, assuming different sets of diagonalization matrices for the leptonic sector, it is easy to convince ourselves that the larger ones come from the doubly charged scalar $Y^{--}$ and vector $U^{--}$ (see Figs.~\ref{fig:MDM_YPP} and \ref{fig:MDM_UPP}). In fact, from the vertices in Eq.~(\ref{aae2}), we can see that the contributions of the neutral scalar are suppressed since they are proportional to the usual charged lepton masses $(\hat{M}^l/v_\rho)O_{\rho1}$. The pseudo-scalar vertex is proportional to $(\hat{M}^l/v_\rho)U_{\rho3}$, thus may be larger than the scalar one but still very suppressed. The vertex of the singly charged scalar $Y^-_1$ is also proportional to $(\hat{M}^l/v_\rho)$, and in this case there are additional suppression factors $\cos\beta V^l_L$, see Eq.~(\ref{aae3 }). Given that the singly charged scalar contribution is negligible, the introduction of the $a_{10}$ term in the scalar potential bring no significant change to our results, but it is necessary for the width decay of the leptons $E^+$. We recall also that in this model the $Z^\prime$ has its coupling with the known charged lepton suppressed, because they are proportional to $\sqrt{1-4s^2_W}$, as can be seen from Eq.~(\ref{aae15}) ($Z'$ is leptophilic in this model).

In the case of the doubly charged scalar $Y^{--}$, it has interactions proportional to $\hat{M}^E/v_\rho>~1$, see Eq.~(\ref{aae4}). Finally, we note that the doubly charged vector bilepton, $U^{--}$ has vertices that are proportional to 
$gV^{l\dagger}_L=(4G_FM^2_W/\sqrt{2})^{1/2}V^{l\dagger}_L$, see Eqs.~(\ref{aae12}) and (\ref{aae13}). Hence, our results also depend on the values of the matrices $V^l_L,$ which we considered.

Most studies of the possibilities of the 3-3-1 models for solving the muon anomaly have considered that the interactions are diagonal in flavor, the scenario we addressed in Sec. \ref{sec:NumericalResultsAMDMsubDiag}. These authors obtain solutions  with low masses for the particles in their models. These interactions are characterized by a coupling strength denoted by $f$ in Ref.~\cite{Jegerlehner:2009ry} from where some authors take off the results for the Feynman diagrams. However, in a particular model, the mass eigenstates appear only after diagonalizing the mass matrices, in doing so the coupling strength $f$ is related to masses of the internal particles in the diagrams and to the unitary matrices that diagonalize the mass matrices. Thus, they are not anymore arbitrary. Moreover, in most works studying $\Delta a_\mu$, usually only one type of exotic particle is considered to address the problem.

To make a comparison, we have calculated the contributions given only by $U^{--}$ and $Y^{--}$ (Figs. \ref{fig:MDM_YPP} and \ref{fig:MDM_UPP}) considering diagonal interactions (i.e. $V_{L,R}^l=1$). When considering only the vector $U^{--}$, although it solves the muon anomaly for $v_\chi$ around 8 TeV, this does not happen for the electron MDM around the same value. From the bottom plot in  the same figure, it seems that the electron MDM can be solved for higher values of $v_\chi$, but these values will not satisfy the muon MDM according to the upper plot. For the case where only $Y^{++}$ contributes, the muon MDM is solved for $v_\chi \approx 100$ GeV while the electron is not solved for this value. So it seems that even with diagonal interaction the MDMs cannot be solved simultaneously.

Of course, our results are in the 331HL model and are not necessarily valid in other 331 models, and also for other values of the matrix $V^l_L$. However they indicate that the analysis in those models should be revisited. Not just because diagonal flavor interactions are the only ones considered, but also because of the effects these exotic particles may have on the electron MDM. The effects on both MDMs can be contradictory, as we have shown in this work.

Once we have determined the value of the VEV $v_\chi$ we can calculate the vector boson masses~\cite{Dias:2006ns} and $m_E$ using:
\begin{eqnarray}
&& M^2_U\approx \frac{\alpha}{4s^2_W}(v^2_\rho+v^2_\chi),\quad  M^2_V\approx \frac{\alpha}{4s^2_W}(v^2_\eta+v^2_\chi), \nonumber \\ &&
M^2_{Z^\prime}=\frac{\alpha^2}{2s^2_W}\frac{(1-2s^2_W)(4+\bar{v}^2_W)+s^4_W(4-\bar{v}^2_W)}{6c^2_W(1-4s^2_W)}, \nonumber \\ &&
\quad m_{E_l}=g_{E_l}\frac{v_\chi}{\sqrt2},
\label{uvmasses}
\end{eqnarray}
where $\bar{v}_W=v_{SM}/v_\chi$, and we see that with $v_\chi=70$ TeV we have $M_U\approx 22$ TeV, $M_V\approx22$ TeV, 
$M_{Z^\prime}\approx 81$ TeV. 

On the other hand, the masses of the scalars $Y^-_{1,2},Y^{--}$ depend also on the dimensionless couplings appearing in the scalar potential, and also on the trilinear term in the scalar potential, $F\,\eta\rho\chi$ denoted by $\alpha$ in  \cite{DeConto:2014fza}. The constant $F$, with dimension of mass, may be small on naturally grounds, or large if it arises from the VEV of a heavy neutral scalar that is singlet under the 3-3-1 symmetry. For this reason the masses of the scalars  $A,Y^-_{1,2},Y^{--}$ and $m_{E_l}$ are used as inputs in our calculations.  On the other hand, the masses of the vectors $Z^\prime,V^-,U^{--}$ depend mainly on $v_\chi$, since the other VEVs are already fixed.

\subsection{The $\mu \rightarrow e \gamma$ decay and $\mu-e$ conversion}

Another possible constraint on the $m_E$ and $v_\chi$ values can come from the $\mu \rightarrow e \gamma$ decay. The diagrams contributing to this process are those in Fig~1 when the intermediate fermions are $E$'s ou neutrinos. This processes have been considered recently in the context of the minimal 3-3-1 model~\cite{Machado:2016jzb} and an early reference is  Ref.~\cite{Liu:1993gy}. Considering only the largest contribution for the branching ratio to this process, we have
\begin{equation}
BR( \mu \rightarrow e \gamma) \propto \frac{54 \alpha}{\pi} \left(
\frac{m_{U}}{m_{Y_1}} \right)^4 \left( \frac{m_E}{m_\mu}
\right)^2 \left(   |(V_Y)_{13}|^2 |(V_Y)_{32}|^2 +  |(V_Y)_{31}|^2
|(V_Y)_{23}|^2 \right)
\end{equation}
where $V_Y=(V^l_R)^T V_L^l$. The actual experimental limit for this is $BR( \mu \rightarrow e \gamma)<0.057 \times 10^{-11}$ \cite{Agashe:2014kda}. Imposing this limit on the above equation, while varying $m_E$ from 0 to 1000 GeV and $v_\chi$ from 10 to 150 TeV, considering all the sets of diagonalization matrices from the sections above, we see that the whole range of values explored is allowed for the diagonal matrices and the first, second and fourth sets of matrices. As for the third set of matrices, none of the values explored for $m_E$ and $v_\chi$ respects the experimental limit for the $\mu \rightarrow e \gamma$ decay branching ratio. Given the simplicity of these results, we decided not to show the respective plots here.

Another potential source of constraints on the parameters of the model is the $\mu-e$ conversion in nuclei. It is easy to see that in the present model such processes do not occur at tree level, since there are no lepton violating neutral currents neither via neutral scalars nor through $Z^\prime$ at this level. Hence, the conversion may happen at least at the 1-loop level. In general, the processes have two types of contributions, the photonic ones in which the photon connects the lepton violating processes occurring at 1-loop level with the nuclei, and the non-photonic ones which, instead of the photon, are mediated by heavy particles and imply four-fermion interactions as $qq\mu e$. Photonic and non-photonic contributions to the $\mu-e$ conversion  involve diagrams similar to those in Fig.~\ref{fig:MDM_todos}.  The photonic contributions deserve more careful study and are beyond the scope of this paper. In the non-photoinc contributions the leptonic loops should interact with quarks by flavor conserving neutral vector ($Z,Z^\prime$) and/or scalars. Hence, these contributions to the $\mu-e$ conversion are already suppressed in the $\mu\to e\gamma$ decay and they do not constraint anymore the parameters of the model.
In this manner, even the rate $7.0\times 10^{-13}$ of this conversion in $^{197}Au$~\cite{Bertl:2006up}  is easily satisfied in the present model.

\section{Conclusions}
\label{sec:con}

In this work we have shown the analytical and numerical results for the electron and muon AMDM at 1-loop level in the 331HL model, comparing them with the experimental results. In the parameter space that has been explored here we have found regions where only the muon AMDM coincides with the experimental value whithin 1$\sigma$, while for the electron case only solutions within 2$\sigma$ and 3$\sigma$ were found. Moreover, even when we found solutions for both electron and muon, the region where they overlapped required very low mass values for the exotic leptons, and this may be a very unlikely scenario. We recall that the masses of vectors $Z^\prime$ and $U^{--}$ are determined mainly by $v_\chi$ once the other VEVs are already fixed, but for the heavy scalars their masses, as for the heavy leptons, also depend on dimensionless parameters.  In this vein, we have used the charged scalars and heavy leptons masses as input in our calculations.

As we said in the previous section, the dominant contributions are due to $Y ^{--}$ and mainly to $U^{--}$. This can be appreciated in Figs. \ref{fig:MDM_YPP} and \ref{fig:MDM_UPP}, respectively. We can see from those figures that, even in these two cases, it is impossible to fit in $1\sigma$ both $\Delta a_e$ and $\Delta a_\mu$. At $3\sigma$ for the electron and less than $2\sigma$ for the muon it is possible to have an agreement with the experimental measurements. However, it leaves the electron case in the same footing as the muon MDM in the SM. For this reason we claim that it is not possible to fit both $\Delta_{e,\mu}$ at the same time. For most of the values explored for the parameters, the contributions of the $a^{331}_e$ increase the value of $\Delta a_e$.    

Other recent works have addressed the muon MDM in different but similar scenarios, in which smaller values for $v_\chi$ and smaller masses for the extra particles were enough to explain the $\Delta a_\mu$~\cite{Kelso:2013zfa,Kelso:2014qka,Ky:2000ku,Binh:2015cba, Taibi:2015ura,Chowdhury:2015rja,Harigaya:2015jba,Ajaib:2015yma,Padley:2015uma,Khalil:2015wua,Abe:2015oca,Chun}. However, none of them have considered at the same time the contributions for the electron MDM from the particles and parameters that solve the muon MDM.

One important difference with other works in 3-3-1 models is that usually the integrals from the Feynmann diagrams were taken from Ref.~ \cite{Jegerlehner:2009ry}. However, the results in the latter paper are quite general and do not take into account details that are model dependent. For instance, they used directly the symmetry eigenstates basis, and in this situation the unitary matrices needed to diagonalize the lepton mass matrices do not appear in the vertices (nor does the VEVs that come from the diagonalization of the scalar mass matrices). Each contribution to the $\Delta a_\mu$ coming from new particles were given just as follows:
\begin{equation}
\Delta a_\mu=\pm\frac{f^2}{4\pi^2}\frac{m^2_\mu}{M^2}I,
\label{jeger}
\end{equation} 
where $M$ is a mass of one of the particle in the loop and $I$ an integral on the Feynman parameter $x$, also depending on parameters as $\epsilon$ and $\lambda$ defined in Sec. \ref{sec:results}. In some of the references above, this integral is simplified, turning into a numerical factor. 

Besides that, in these works all vertices are reduced to the exotic particle with mass $M$, coupling to muons with coupling strength $f$. In our case, the mass eigenstates particles in the diagrams are obtained from the symmetry eigenstates and the unitary matrices relating both basis appear in the vertices. Here these matrices are incorporated in the factors $S,P$ when the fields are (pseudo)scalars and $V$ and $A$ when they are the vectors.

For instance, in Ref.~\cite{Kelso:2013zfa} the masses of all internal fermions in the loops are neglected in the reduced m331 model. This could be done because in this model the only internal fermions are the known leptons. In the m331 model (reduced or not)  the interactions with the doubly charged scalar or vector boson have non-diagonal vector and axial-vector interactions that have not been considered in most works mentioned above. Moreover, in \cite{Kelso:2013zfa} the only mixing matrix in the lepton sector is the PMNS. The effect of this matrix is nullified by the small neutrino masses and the unitarity of the PMNS matrix. On the other hand, in the minimal 3-3-1 the charged current coupled to $U^{++}$ is
\begin{equation}
\mathcal{L}^{cc}_{ll}=-i\frac{g}{2\sqrt2}  
\bar{l^c}\gamma^\mu[(V_U-V^T_U)-\gamma_5(V_U+V^T_U)]lU^{++}_\mu+H.c.
\label{uu}
\end{equation}
where $l=(e,\mu,\tau)^T $ and $V_U=(V^l_R)^TV^l_L$. We see that there are non-diagonal vector currents and the unitary matrices are such that $V^{l\dagger}_LM^lV^l_R=\hat{M}^l=\textrm{diag}(m_e,m_\mu,m_\tau)$, it is an extreme fine tuning to assume that $V_U$ is the unit matrix.  Hence, in the minimal 331 model it is not advisable to neglect the lepton mixing.

In the model explored in this work this type of current involves $E$ and $l$, and $V^l_L$, see Eq.~(\ref{aae12}).
For instance, in \cite{deS.Pires:2001da} it is shown that the doubly charged scalar may solve the muon anomaly, using scalars with  masses  around  hundreds of GeV's. The scheme in \cite{deS.Pires:2001da} could, at first sight, be realized in the m331. It is so because in that model, a singlet doubly charged scalar (under the SM symmetries) belonging to  sextet couples only with right-handed charged leptons, i.e., the respective interactions are $\overline{(l_R)^c}V^{lT}_LG^SV^l_Rl_R$, where $G^S$ is the Yukawa matrix in the interactions of leptons with the scalar sextet. Moreover, $G^S$ is not proportional to the charged lepton masses because they have another source of mass: the interactions with the triplet $\eta$. Hence, the arbitrary matrix $V^{lT}_LG^SV^l_R$ has to be taken into account in the m331. 

Another example, in Ref.~\cite{Kelso:2014qka}, were considered solutions to the muon anomaly in the context of the minimal 331, the 331 model with heavy neutral fermions, and the 331 model with heavy charged leptons but with five left-handed lepton triplets, thus the latter one is different from the 331HL model here considered. Again, the results are all approximated and since all flavor mixings are neglected their results are, as before, of the form $\Delta a_\mu\propto m^2_\mu/M^2$. The masses of the usual charged leptons can be neglected in the loops because they appear as $\epsilon_\mu=m_\nu/m_\mu$ or $\lambda_l=m_\mu/M$. The first is negligible by the neutrino masses and the second by the value of $M$. Another important point to be stressed is that, in \cite{Kelso:2014qka,Ky:2000ku} not all the particles in each model are considered. 
Moreover, it happens that in the model they were considering, the importance of the scalar contributions cannot be neglected \textit{a priori} because they depend on the Yukawa couplings through the unitary matrices that diagonalize quarks and lepton mass matrices, as discussed in the previous paragraph. For instance, in the m331 there are FCNC via the scalar sector, and in this case the scalar contributions may be as important as in the quark sector~\cite{Machado:2013jca}. 

We must put in context our results. On one hand, they were obtained in the 331HL where there are no flavor changing neutral interactions in the lepton sector. Moreover, unlike in the m331, the singly charged vector bilepton $V^+$ does not contribute to the MDM of the known charged leptons. For example, this is not the case with the m331, where there are more scalar multiplets and FCNC in the lepton sector via the exchange of neutral Higgs. On the other hand, they were obtained using the matrices $V_L^l$ in in Sec. \ref{sec:NumericalResultsAMDM}. The matrices $V^l_R$ which we presented are possible solutions, they may have other values and it is possible that these values might imply solutions for both MDMs. The moral of the story, is that besides the unitary matrices, the MDM of the electron must be taken into account when solutions for the case of muon are proposed, once both MDMs may be incompatible. Otherwise, the results cannot be considered definitive because they depends also on the solutions for the matrices $V^l_{L,R}$ and different solutions imply different lower bounds on the phenomenology of the model~\cite{Machado:2016jzb}.

\acknowledgments
The authors would like to thank CNPq for full (G.D.C.) and partial (V.P.) financial support. 

\newpage 
\appendix

\section{Scalar Potential}
\label{sec:potential}

The most general scalar potential is given by
\begin{eqnarray}
%\begin{split}
V\left(\chi,\eta,\rho\right)&=& \,\mu_{1}^{2}\chi^{\dagger}\chi+\mu_{2}^{2}\eta^{\dagger}\eta+ \mu_{3}^{2}\rho^{\dagger}\rho+\left( \alpha\, \epsilon_{ijk}\chi_{i}\rho_{j}\eta_{k}+H.c. \right)+a_{1}\left(\chi^{\dagger}\chi\right)^{2} \nonumber \\ &+& a_{2}\left( \eta^{\dagger}\eta \right)^{2} +a_{3}\left( \rho^{\dagger}\rho \right)^{2}+a_{4}\left(\chi^{\dagger}\chi\right)\left( \eta^{\dagger}\eta \right)+a_{5}\left(\chi^{\dagger}\chi\right)\left( \rho^{\dagger}\rho \right) \nonumber \\ & +& a_{6}\left( \rho^{\dagger}\rho \right)\left( \eta^{\dagger}\eta \right)  +a_{7}\left(\chi^{\dagger}\eta\right)\left(\eta^{\dagger}\chi\right)+a_{8}\left(\chi^{\dagger}\rho\right)\left(\rho^{\dagger}\chi\right)\nonumber \\ &+& a_{9}\left(\rho^{\dagger}\eta\right)\left(\eta^{\dagger}\rho\right) + a_{10}[ (\chi^\dagger \eta )(\rho^\dagger \eta)+(\eta^\dagger \chi)(\eta^\dagger \rho)]
%\end{split}
\label{scalarpotencial}
\end{eqnarray}
we have assumed $a_{10}$ real. The scalar potential with $a_{10}=0$ has been considered for instance in Ref.~\cite{Montero:1998yw,DeConto:2014fza}. If this term is not zero the only sector which is modified is that of the singly charged scalars. In the basis $(\eta^-_1,\,\rho^-,\,\eta^-_2,\,\chi^-)$ the mass matrix becomes
\begin{equation}
M^2=\left(
\begin{array}{cccc}
\frac{ a_9  v_\rho^2}{2}+\frac{v_\chi \alpha  v_\rho}{\sqrt{2} v_\eta} & \frac{ a_9  v_\eta v_\rho}{2}+\frac{v_\chi \alpha }{\sqrt{2}} & \frac{ a_{10}  v_\rho v_\chi}{2} & \frac{ a_{10}  v_\eta v_\rho}{2} \\
 & \frac{ a_9  v_\eta^2}{2}+\frac{v_\chi \alpha  v_\eta}{\sqrt{2} v_\rho} & \frac{ a_{10}  v_\eta v_\chi}{2} & \frac{ a_{10}  v_\eta^2}{2} \\
 & & \frac{ a_7  v_\chi^2}{2}+\frac{v_\rho \alpha  v_\chi}{\sqrt{2} v_\eta} & \frac{ a_7  v_\eta v_\chi}{2}+\frac{v_\rho \alpha }{\sqrt{2}} \\
 &  &  & \frac{ a_7  v_\eta^2}{2}+\frac{v_\rho \alpha  v_\eta}{\sqrt{2} v_\chi} \\
\end{array}
\right).
\label{salva}
\end{equation}
The symmetry eigenstates are related to the mass eigenstates $h^-_i=Y^-_1,\, G^-_1,\,Y^-_2,\,G^-_2$ by the orthogonal  matrix $U$ which we will not write explicitly here. We just write the symmetry eigenstates $x^+=\eta^+_1,\rho^-,\eta^+_2,\chi^+$, in term of the mass eigenstates as follows: $x^+_a=\sum U_{ai}h^-_i$. This mass matrix has two null eigenvalues, corresponding to two Goldstone bosons, and two non-null eigenvalues corresponding to the charged scalars $Y_1^+$ and $Y_2^+$. Notice that if $a_{10}=0$, the scalars $\eta^-_1,\rho^-$ do not mix with  $\eta^-_2,\chi^-$.

\section{Interactions}
\label{sec:interactions}

\subsection{Lepton-scalar vertices}
\label{subsec:leptonscalars}

In the following equations we denote: 
\begin{eqnarray} 
\hat{M}^l=\textrm{diag}
(m_e,m_\mu,m_\tau),\quad  M^E=diag(m_{E_e},m_{E_\mu},m_{E_\tau}),  
\label{aae1}
\end{eqnarray} 
with $\nu=(\nu_e,\nu_\mu,\nu_\tau), l=(e,\mu,\tau), E=(E_e,E_\mu,E_\tau)$. 
The neutral (pseudo)scalars $\bar{l}(\sqrt{2}\hat{M}^l/v_\rho)l\cdot\sum^3_{i=1}O_{\rho i}h_i$, and   
$\bar{l}(\hat{M}^l/v_\rho)\gamma_5l\cdot\sum^3_{i=1}U_{\rho i}A_i$. Since in this model there is only one physical pseudo-scalar we denote $A_3\equiv A^0$.  When $h_1$ is the scalar with mass about 125 GeV, with $O_{\rho1}=0.42$, see ~\cite{Machado:2013jca}. Here we have the form factors appearing in (\ref{as}) ($i$ fixed) 
\begin{eqnarray}
S^l_{h_i}=\frac{\sqrt{2}\hat{M}^l}{v_\rho}O_{\rho i},\quad
P^l_A= \frac{\sqrt{2}}{v_\rho^2\sqrt{\frac{1}{v_\chi^2}^+\frac{1}{v_\eta^2}+\frac{1}{v_\rho^2}}}\hat{M}^lU_{\rho 3}.
\label{aae2}
\end{eqnarray}
Notice that the scalar $h_1$ has no $P$ vertex and the pseudo-scalar $A^0$ has no $S$ vertex because in this model there are no FCNCs in the Higgs sector. Hence, they cancel out. This is not the case in the minimal 331 model since in this case there are FCNC in the Higgs sector and the vertices are chiral. 

From the Yukawa interactions in Eq.~(\ref{yukawa}) we obtain the following vertices. For the singly charged scalar $Y^+_1$, the vertex  $\bar{\nu}_{iL}E_{jR}Y_1^+$ implies
\begin{eqnarray}
S^l_{Y_1(Y_2)}=P^l_{Y_1(Y_2)}=U_1(U_2)\frac{\sqrt{2}}{v_\rho} (V^{l\dagger}_L)_{ij}\hat{M}^l,
\label{aae3 }
\end{eqnarray}
where $U_1(U_2)$ represent the projection of a given singly charged scalar into $Y_{1,2}$.
%where $\cos\beta=v_\eta/\sqrt{v^2_\eta+v^2_\rho}$ and $\sin\beta=v_\rho/\sqrt{v^2_\eta+v^2_\rho}$. 

Finally, for the doubly charged scalar we have $\bar{l}EY^{--} \rightarrow i(S_Y +\gamma_5P_{Y})$, with
\begin{eqnarray}
&&
S_{Y^{--}}= \frac{\sqrt2}{v_\rho} [ \cos\alpha  \hat{M}^E V_L^{l\dagger}+ 
\sin\alpha V^{l\dagger}_L\hat{M}^l],
\nonumber \\&&
P^l_{Y^{--}}= \frac{\sqrt2}{v_\rho}[ -\cos\alpha \hat{M}^EV^{l\dagger}_L+\sin\alpha 
V_L^{l\dagger}\hat{M}^l],
\label{aae4} 
\end{eqnarray}
where $\cos\alpha=v_\rho/\sqrt{v^2_\rho+v^2_\chi}$ and $\sin\alpha=v_\chi/\sqrt{v^2_\rho+v^2_\chi}$.
where the constants $V^l_S$ and $A^l_S$ are defined in Sec.~\ref{sec:model}. We assumed that the heavy leptons are in the diagonal basis, $M^E=\hat{M}^E$ and that in (\ref{aae4}), the $v_\chi$ is real, i.e., $\theta_\chi=0$. 

The numerical values for the $V^l_{L,R}$ matrices were obtained in Ref.~\cite{DeConto:2014fza} and \cite{Machado:2016jzb}. The different sets os matrices considered can be seen in Sec. \ref{sec:NumericalResultsAMDM}.

In the interactions above, the matrix $V^l_R$ and the matrix for the Yukawa couplings can always be eliminated by using the mass matrix. This behavior is typical of the present model in which there is no flavor changing neutral currents. In the m331 one, $V^l_R$ survive in some interactions with scalars. Again, this is a consequence that in the 331HL model there are no FCNCs in the scalar sector.

The extra factors in the interactions above come from the projection on the SM-like scalars, see~\cite{DeConto:2014fza} for details.

\subsection{Scalar-photon vertices} 
\label{subsec:vertices_gauge_escalar}

From the lepton Lagrangian we may find the interaction with photon of the known leptons, $l(e\,\mu\,\tau)^T$  and extra leptons $E=(E_e\,E_\mu\,E_\tau)^T$:
\begin{equation}
\mathcal{L}_I= (-e\bar{l}\gamma_\mu l+e\bar{E}\gamma_\mu E) A^\mu
\label{aae5}
\end{equation}
and identify the electron charge as
\begin{equation}
e=\frac{gt}{\sqrt{1+4t^2}}=gs_W,
\label{aae6}
\end{equation}
where $e$ is the modulus of the electron charge and $t^2=s^2_W/(1-4s^2_W)$.

Now, from the covariant derivatives of the scalar's Lagrangian we can find the vertices for the interactions between scalars and photons:
\begin{eqnarray}
&& A_\mu Y^{++}Y^{--}\rightarrow i2e(k^--k^+)_\mu,\quad
A_\mu Y_2^+Y_2^-\rightarrow ie(k^--k^+)_\mu,\nonumber \\ && 
A_\mu Y_1^+Y_1^-\rightarrow ie(k^--k^+)_\mu.
\label{aae7}
\end{eqnarray}
The terms $k^{+}$ and $k^-$ 
indicate, respectively, the momenta of the positive and negative charge scalars, and both should be considered incoming into the vertex. 

\subsection{Gauge vertices} 
\label{subsec:gaugevertices}

For three gauge bosons denoted generically by $X,Y$  and $Z$, with all momenta incoming (denoted by $k$), the vertex is:
\begin{equation}
X_\nu Y_\lambda Z_\mu \rightarrow i G_{XYZ} \left[g_{\nu\lambda}\left(k^X-k^Y\right)_\mu+g_{\lambda\mu}\left(k^Y-k^Z\right)_\nu
+g_{\mu\nu}\left(k^Z-k^X\right)_\lambda\right]
\label{aae8} 
\end{equation}
The proportionality constants are:
\begin{eqnarray}
&& G_{WWA}=e,\quad
G_{VVA}=-e,\quad
G_{UUA}=-2e,\quad
G_{WWZ}=-e/t_W,\nonumber \\&&
G_{WWZ'}=0,
G_{VVZ}=-\frac{e}{2}\left(\frac{1}{t_W}+3t_W\right),\quad
G_{VVZ'}=e\frac{\sqrt{3}}{2}\frac{\sqrt{1-4s_W^2}}{s_Wc_W},\nonumber \\&&
G_{UUZ}=\frac{e}{2}\left(\frac{1}{t_W}-3t_W\right),
G_{UUZ'}=e\frac{\sqrt{3}}{2}\frac{\sqrt{1-4s_W^2}}{s_Wc_W}.
\label{aae9}
\end{eqnarray}
where $s_W$, $c_W$ and $t_W$ are, respectively, the sine, cosine and tangent of the weak mixing angle, $\theta_W$.

\subsection{Charged gauge-lepton interactions}
\label{subsec:glint}

The Lagrangian terms for interactions among charged gauge bosons and leptons may be written as follows:
\begin{itemize}
\item $\nu$-type and l-type leptons:
\begin{equation}
\mathcal{L}_{\nu l}=\frac{g}{2\sqrt{2}}\bar{\nu}_i\gamma^\mu\left(1-\gamma_5\right)(V_{PMNS})_{ij}l_{j}W^{+}_\mu + H.c.
\label{aae10}
\end{equation}
for $i=1,2,3$ and $j=1,2,3$ and $V_{PMNS}$ is the Pontecorvo-Maki-Nagaoka-Sakata matrix. We use the values from~\cite{GonzalezGarcia:2012sz}. We note that because of the small neutrino masses the unitarity of the matrix $V_{PMNS}$ erases the effects of the lepton mixing. Since the neutrino masses are rather small, the non-digonal interactions in (\ref{aae10}) vanish because $V_{PMNS}$ is an unitary matrix. 

\item $E$-type and $\nu$-type leptons:
\begin{equation}
\mathcal{L}_{E\nu}=\frac{g}{2\sqrt{2}}\bar{E}_i \gamma^\mu\left(1-\gamma_5\right)(U^\nu_L)_{ij}\,\nu_j\,V^{+}_\mu + H.c.
\label{aae11}
\end{equation}
for $i=1,2,3$. $U^\nu_L$ is the unitary matrix relating the active neutrino symmetry eigenstates with the mass eigenstates. We also assume that the heavy leptons are in the diagonal basis.
 
\item $E$-type and $l$-type leptons:
\begin{equation}
\mathcal{L}_{El}=\frac{g}{2\sqrt{2}}\bar{E}_i \gamma^\mu\left(1-\gamma_5\right) (V_L^{l\dagger})_{ij} l_{j}U^{++}_\mu + H.c.
\label{aae12}
\end{equation}
for $i,j=1,2,3$ and $V^l_L$ is given in Sec. \ref{sec:NumericalResultsAMDM}. Notice that in this model
\begin{equation}
V^l_U=A^l_U=\frac{g}{2\sqrt2}V^{l\dagger}_L,
\label{aae13}
\end{equation}.

\end{itemize}

\subsection{Neutral gauge-lepton interactions} 
\label{subsec:geugeleptoninteractions}

Considering the following Lagrangian for the neutral interactions:
\begin{eqnarray}
\mathcal{L}_{NC}=-\frac{g}{2c_W}\sum_i \bar{\psi}_i\gamma^\mu \left[\left(g_V^i-g_A^i \gamma_5\right)Z_\mu+\left(f_V^i-f_A^i \gamma_5\right)Z'_\mu \right]\psi_i
\label{aae14}
\end{eqnarray}
where $\psi_i$ can be any lepton mass eigenstate, we find the $g_V^i$, $g_A^i$, $f_V^i$ and $f_A^i$ to be:
\begin{eqnarray}
&& g_V^\nu=\frac{1}{2}, \quad g_A^\nu=\frac{1}{2},\quad 
g_V^l=-\frac{1}{2}+s_W^2,\quad
g_A^l=-\frac{1}{2}+s_W^2,\nonumber \\&&
g_V^E=g_A^E=-s_W^2,
\label{aae15} 
\end{eqnarray}

\begin{eqnarray}
&& 
f_V^\nu=-\frac{\sqrt{1-4s_W^2}}{2\sqrt{3}},\quad
f_A^\nu=-\frac{\sqrt{1-4s_W^2}}{2\sqrt{3}},\nonumber \\&&
f_V^l=-\frac{\sqrt{1-4s_W^2}}{2\sqrt{3}},\quad
f_A^l=-\frac{\sqrt{1-4s_W^2}}{2\sqrt{3}},\nonumber \\&&
f_V^E=\frac{\sqrt{1-4s_W^2}}{2\sqrt{3}},\quad
f_A^E=\frac{\sqrt{1-4s_W^2}}{2\sqrt{3}}
\label{aae16}
\end{eqnarray}

The extra charged leptons are vectorial with respect to $Z$ and $Z^\prime$. Notice also that the $Z^\prime$ is leptophobic with respect to the usual charged leptons.

\newpage


\end{document}